\documentclass[structabstract]{aa}  

\usepackage{graphicx}
\usepackage{txfonts}
\usepackage{natbib}
\usepackage{color}
\usepackage{threeparttable}

\newcommand{\htwo}{H$_{2}$}

\newcommand{\htwoo}{H$_{2}$O}

\newcommand{\msol}{$M_{\odot}$}

\newcommand{\mdot}{$\dot{M}$}

\newcommand{\lsol}{$L_{\odot}$}
{

\newcommand{\tkin}{$T_{\rm kin}$}
\newcommand{\tdust}{$T_{\rm dust}$}
\newcommand{\ngas}{$n_{\rm gas}$}

\newcommand{\tastar}{$T_{\rm A}^{\star}$}
\newcommand{\tmb}{$T_{\rm mb}$}

\newcommand{\teff}{$T_{\rm eff}$}
\newcommand{\kms}{km$\,$s$^{-1}$}

\newcommand{\chgas}{[C]/[H]$_{\rm gas}$}
\newcommand{\ohgas}{[O]/[H]$_{\rm gas}$}

\newcommand{\gdrat}{$\Delta_{\rm gas/dust}$}

\newcommand{\cplus}{C$^{+}$}
\newcommand{\catom}{C$^{0}$}
\newcommand{\ci}{[\ion{C}{i}]}
\newcommand{\cii}{[\ion{C}{ii}]}
\newcommand{\cosixfive}{CO$\,6$--$5$}
\newcommand{\cionezero}{[\ion{C}{i}]$\,1$--$0$}
\newcommand{\citwoone}{[\ion{C}{i}]$\,2$--$1$}
\newcommand{\ciiline}{[\ion{C}{ii}]$\,158\,\mu$m}

\newcommand{\herschel}{\emph{Herschel}}
\newcommand{\hso}{\emph{Herschel}~Space~Observatory}

\newcommand{\champp}{CHAMP$^{+}$}
\newcommand{\noobs}{$\ldots$}

\begin{document}

\title{Observations and modelling of CO and [\ion{C}{i}] in protoplanetary disks}
\subtitle{First detections of [\ion{C}{i}] and constraints on the carbon abundance}
\titlerunning{A survey of CO and [\ion{C}{i}] in protoplanetary disks}

   \author{
   M. Kama\inst{1}
          \and
          S. Bruderer\inst{2}
          \and
          M. Carney\inst{1}
          \and
          M. Hogerheijde\inst{1}
          \and
          E.F. van Dishoeck\inst{1}
          \and
          D. Fedele\inst{2}
          \and
          A. Baryshev\inst{3,4}
          \and
          W. Boland\inst{1,5}
          \and
          R. G\"{u}sten\inst{6}
          \and
          A. Aikutalp\inst{4}
          \and
          Y. Choi\inst{4}
          \and
          A. Endo\inst{7}
          \and
          W. Frieswijk\inst{1,8}
          \and
          A. Karska\inst{2,9}
          \and
          P. Klaassen\inst{1,10}
          \and
          E. Koumpia\inst{4}
          \and
          L. Kristensen\inst{1,11}
          \and
          S. Leurini	\inst{6}
          \and
          Z. Nagy\inst{4,12}
          \and
          J.-P. Perez Beaupuits\inst{6}
          \and
          C. Risacher\inst{3,6}
          \and
          N. van der Marel\inst{1}
          \and
          T.A. van Kempen\inst{1}
          \and
          R.J. van Weeren\inst{1,11}
          \and
          F. Wyrowski\inst{6}
          \and
          U.A. Y{\i}ld{\i}z\inst{1,13}
          }

\institute{
		Leiden Observatory, P.O. Box 9513, NL-2300 RA, Leiden, The Netherlands, \email{mkama@strw.leidenuniv.nl}
		\and
		Max Planck Institut f\"{u}r Extraterrestrische Physik, Giessenbachstrasse 1, 85748 Garching, Germany
		\and
		SRON Netherlands Institute for Space Research
		\and
		Kapteyn Astronomical Institute, P.O. Box 800, 9700 AV Groningen, The Netherlands
		\and
		NOVA, J.H. Oort Building, P.O. Box 9513, 2300 RA Leiden, The Netherlands
		\and
		Max-Planck-Institut f\"{u}r Radioastronomie, Auf dem H\"{u}gel 69, 53121, Bonn, Germany
		\and
		Kavli Institute of Nanoscience, Delft University of Technology, Lorentzweg 1, 2628 CJ Delft, The Netherlands
		\and
		ASTRON, the Netherlands Institute for Radio Astronomy, Postbus 2, 7990 AA, Dwingeloo, The Netherlands
		\and
		Astronomical Observatory Institute, Faculty of Physics, A. Mickiewicz University, Sloneczna 36, 60-286, Poznan, Poland
		\and
		UK Astronomy Technology Center, Royal Observatory Edinburgh, Blackford Hill, Edinburgh EH9 3HJ, UK
		\and
		Harvard-Smithsonian Center for Astrophysics, 60 Garden Street, Cambridge, MA 02138, USA
		\and
		Department of Physics and Astronomy, University of Toledo, 2801 West Bancroft Street, Toledo, OH 43606, USA
		\and
		Jet Propulsion Laboratory, California Institute of Technology, 4800 Oak Grove Drive, Pasadena, CA 91109, USA
             }

   \date{}

 
  \abstract
   {The gas-solid budget of carbon in protoplanetary disks is related to the composition of the cores and atmospheres of the planets forming in them. The key gas-phase carbon carriers CO, C$^{0}$ and C$^{+}$ can now be observed regularly in disks.}
   {The gas-phase carbon abundance in disks has thus far not been well characterized observationally. We aim to obtain new constraints on the [C]/[H] ratio in a large sample of disks, and to compile an overview of the strength of [\ion{C}{i}] and warm CO emission.}
   {We carried out a survey of the CO$\,6$--$5$ and [\ion{C}{i}]$\,1$--$0$ and $2$--$1$ lines towards $37$ disks with the APEX telescope, and supplemented it with [\ion{C}{ii}] data from the literature. The data are interpreted using a grid of models produced with the DALI disk code. We also investigate how well the gas-phase carbon abundance can be determined in light of parameter uncertainties.}
   {The CO$\,6$--$5$ line is detected in $13$ out of $33$ sources, the [\ion{C}{i}]$\,1$--$0$ in $6$ out of $12$, and the [\ion{C}{i}]$\,2$--$1$ in $1$ out of $33$. With separate deep integrations, the first unambiguous detections of the [\ion{C}{i}]~$1$--$0$ line in disks are obtained, in TW~Hya and HD~100546. }
   {Gas-phase carbon abundance reductions of a factor $5$--$10$ or more can be identified robustly based on CO and [\ion{C}{i}] detections, assuming reasonable constraints on other parameters. The atomic carbon detection towards TW~Hya confirms a factor $100$ reduction of [C]/[H]$_{\rm gas}$ in that disk, while the data are consistent with an ISM-like carbon abundance for HD~100546. In addition, BP~Tau, T~Cha, HD~139614, HD~141569, and HD~100453 are either carbon-depleted or gas-poor disks. The low [\ion{C}{i}]~$2$--$1$ detection rates in the survey mostly reflect insufficient sensitivity to detect T~Tauri disks. The Herbig~Ae/Be disks with CO and [\ion{C}{ii}] upper limits below the models are debris disk like systems. A roughly order of magnitude increase in sensitivity compared to our survey is required to obtain useful constraints on the gas-phase [C]/[H] ratio in most of the targeted systems.}
   \keywords{protoplanetary disks; surveys; submillimeter: planetary systems}
   \maketitle

\section{Introduction}

Carbon is one of the most abundant elements in the Universe, and is central to interstellar and terrestrial chemistry, and to planetary climate  \citep{HenningSalama1998, Unterbornetal2014}. The carbon content of planets is determined by chemical and physical processes before and during the protoplanetary disk stage. To elucidate this stage-setting for planetary compositions, we present a survey of carbon reservoirs in $37$ protoplanetary disk systems, including the first firm detections of sub-millimetre atomic carbon lines from disks.

The carbon budget in inter- and circumstellar material broadly consists of refractory -- e.g., graphite or amorphous carbon -- and volatile material -- atoms, simple and complex molecules and ices. There is evidence for a rapid recycling between these \citep{Jones2014}. The elemental abundance of gas-phase carbon with respect to hydrogen, \chgas, in the interstellar medium is $(1-2)\times 10^{-4}$ \citep{Cardellietal1996, Parvathietal2012}, while the solar value is $2.69\times 10^{-4}$ \citep{Asplundetal2009}. This implies that, in the ISM, volatile and refractory reservoirs each contain about $50$\% of elemental carbon. A smaller fraction of interstellar carbon atoms, up to $\lesssim 5$\%, are bound in polycyclic aromatic hydrocarbon molecules \citep[PAHs,][]{Tielens2008}.

In the surface layers of protoplanetary disks, with increasing shielding from (inter)stellar ultraviolet photons \cplus, \catom\ and CO are the dominant gas-phase carbon carriers. Depending on the ionization state, chemical history and gas temperature, CO$_{2}$ and small hydrocarbons may carry large fractions of the volatile carbon \citep[e.g.,][]{Berginetal2014FD, Pontoppidanetal2014FD}. At low dust temperatures (below $\sim 25\,$K for CO), the molecular carriers form icy layers on dust, where they can be further processed into complex organics which may evaporate when brought into warmer conditions. Vertical and radial mixing may give rise to a flow of carbon from the warm, tenuous gas in the disk atmosphere into cold, icy reservoirs which evolve and migrate into the inner disk, where volatile and perhaps even refractory carbon is channeled into the gas phase \citep[e.g.,][]{Leeetal2010, Pontoppidanetal2014PPVI}. The co-evolution of the various reservoirs is reflected in the gas-phase elemental abundance of carbon in the outer disk atmosphere, where volatiles likely cannot return to once locked in large icy grains and transported to the inner disk.

Observational estimates of the carbon budget in disks are difficult. The gas phase is the most accessible, with CO being the
dominant reservoir in molecular gas. However, the disk-averaged CO abundance can be much less than the canonical value of CO/\htwo~$\approx 10^{-4}$ due to the abovementioned freezeout and due to photodissociation in the upper layers
\citep[e.g.,][]{vanZadelhoffetal2001,Dutreyetal2003,Chapillonetal2010}. This, combined with optical depth effects, makes recovering the elemental \chgas\ from CO alone tricky.

Neutral and ionized atomic carbon, \catom\ and \cplus\ (noted \ci\ and \cii\ where line emission is concerned), consecutively become the main gaseous carbon reservoirs in the UV-irradiated surface layers of the disk. However, \cii\ cannot be observed from the ground and \hso\ data contain emission from residual envelope material around disks \citep{Fedeleetal2013a, Fedeleetal2013b, Dentetal2013}. Neutral atomic carbon may also have a non-disk emission component, but this is easier to check as the observations are resolved in velocity and additional pointings can be taken. It is thus a promising tracer of the carbon abundance in disk atmospheres, but its disk contribution has not yet been unambiguously detected \citep{Chapillonetal2010, Panicetal2010, Casassusetal2013, vanderWieletal2014, Tsukagoshietal2015}.

\begin{figure*}[!ht]
\includegraphics[clip=, width=1.0\linewidth]{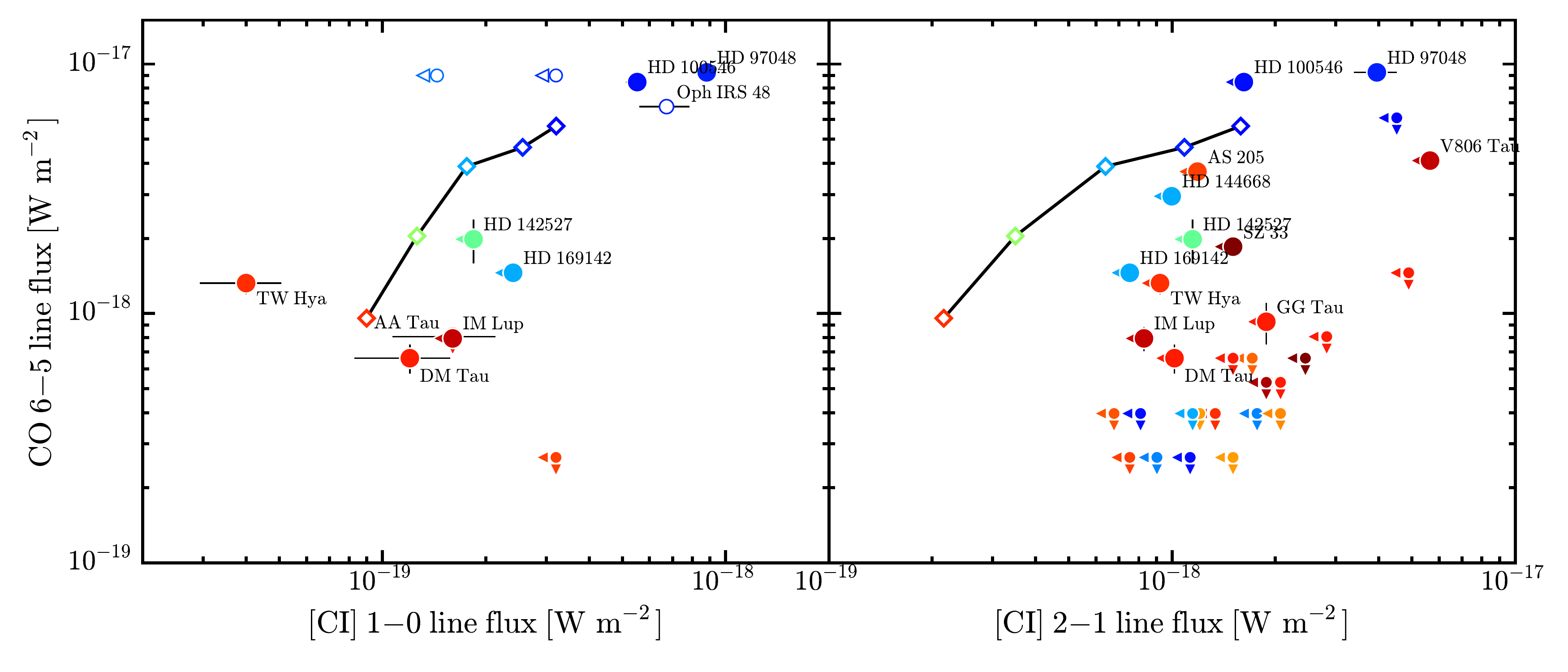}
\caption{Fluxes and upper limits (at $3\sigma$ confidence) from our APEX observations of the \cionezero\ and $2$--$1$ and the \cosixfive\ transitions. Large circles indicate sources with a detection on at least one axis. Empty symbols mark sources for which only the \cionezero\ line was observed. \cosixfive\ and \citwoone\ were always observed in parallel. Colours show the stellar effective temperature, with late-type stars red and early-type stars blue. The black line connects fiducial disk models from Section~\ref{sec:modelling} for the stellar spectral type range of our observations (diamonds from red to blue, \teff~$\in{[4000,12000]}\,$K in steps of $2000\,$K, assumed distance $140\,$pc).}
\label{fig:mainplot3}
\end{figure*}

Through comprehensive modelling, the total gas-phase abundance of carbon was found to be depleted by a factor of $2$ to $10$ with respect to an adopted volatile carbon abundance of \chgas~$=2.4\times10^{-4}$ in the HD~100546 disk \citep{Brudereretal2012}. For the T~Tauri disk system TW~Hya, \citet{Favreetal2013} inferred a deficiency of up to two orders of magnitude in carbon abundance, based on C$^{18}$O observations and a thus far unique bulk gas mass measurement via HD \citep{Berginetal2013}. However, the C$^{18}$O-to-H$_{2}$ conversion may be impacted somewhat by isotopolog-selective CO photodissociation \citep{Miotelloetal2014}.

We observed \ci\ towards a large number of disks using the \emph{Atacama Pathfinder EXperiment} (APEX) telescope at Cerro Chajnantor. We also present deep follow-up observations which resulted in disk detections. In our analysis, we focus on the disk atmosphere, where physical-chemical models allow to relate observations of \ci\ and CO emission to the total gas-phase elemental carbon abundance.

\begin{figure*}[!ht]
\includegraphics[clip=, width=1.0\linewidth]{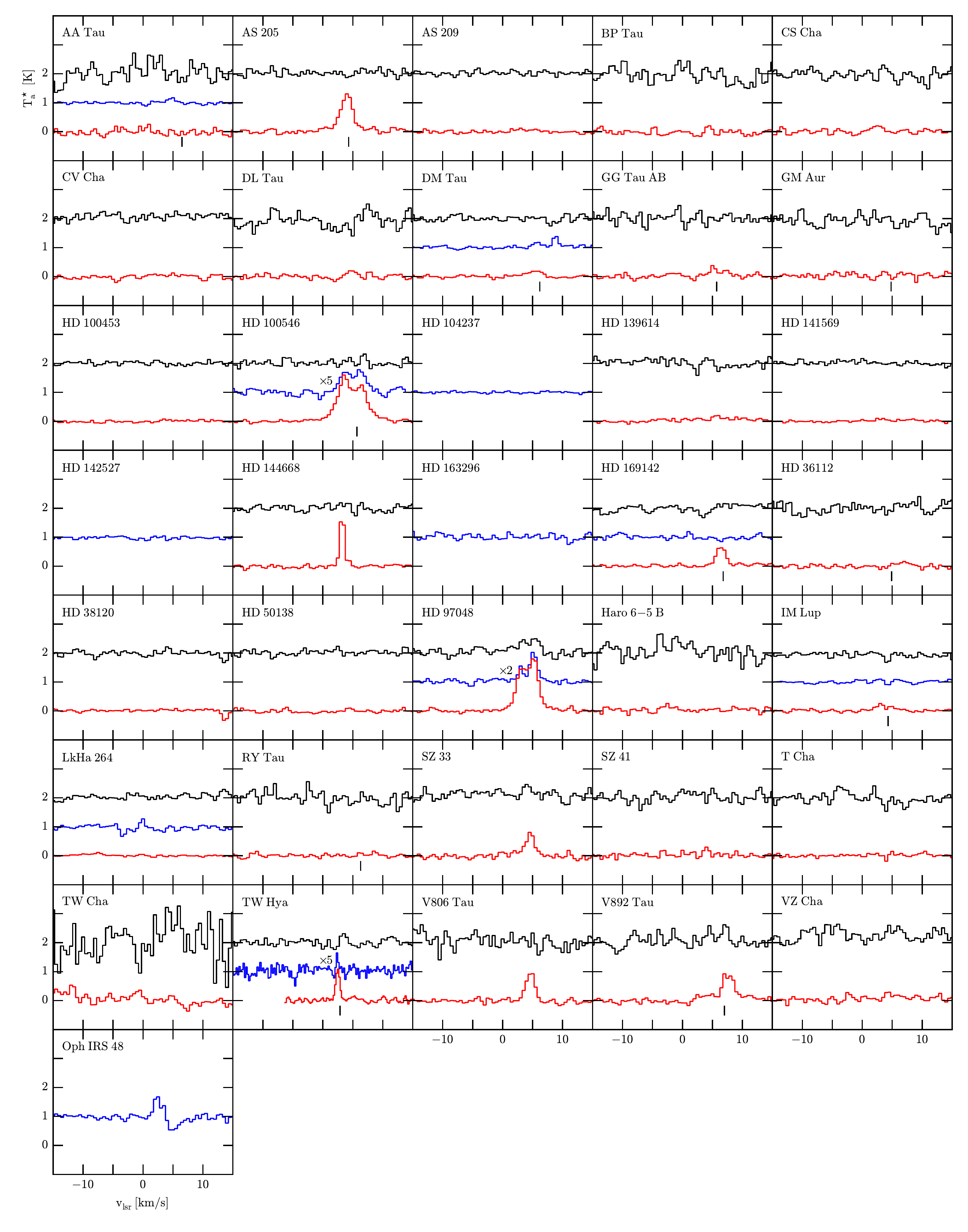}
\caption{All observed spectra, rebinned to a resolution of $0.5$~\kms. From bottom to top in each panel, the spectra are \cosixfive\ (red), \cionezero\ (blue) and \citwoone\ (black). The measured fluxes and noise levels are reported in Table~\ref{tab:sources}. Vertical offsets of $1\,$K have been applied for clarity.}
\label{fig:allspec}
\end{figure*}

\section{Observations}\label{sec:obs}

\subsection{Sample selection}
The disks targeted in our survey (Table~\ref{tab:sources}) are some of the closest and best studied ones. The main selection criteria were observability with APEX, the availability of supplementary data, and proximity to the Solar System. The spectral types range from late-B (HD~141569) to mid-M (Haro~6-5B, Sz~33). 

Studies of disks with single dish instruments are complicated because the large beam can capture extended emission from a surrounding cloud. The current sample includes a number of sources for which previous single dish observations of CO and other species have demonstrated that most of the emission originates from the disk \citep[e.g.,][]{Thietal2004, vanKempenetal2007, PanicHogerheijde2009b, Kastneretal2008, Zuckermanetal1995}. The observed sample includes protoplanetary disks with no known inner hole (e.g., HD~163296) as well as transitional disks with inner holes of up to several tens of au in radius (e.g., HD~100546, HD~169142). Since the disk emission component is dominated by the outer disk, which has similar characteristics in both types of sources, the presence or absence of such holes is of little importance for the purposes of this study. After the initial shallow survey, HD~97048, HD~100546 and TW~Hya were chosen for deep follow-up integrations.

\subsection{APEX observations}
The shallow survey observations of $37$ disks were carried out with the $2{\times}7$-pixel \champp\ \citep{Kasemannetal2006} and the single-pixel FLASH \citep{Heymincketal2006} instruments on APEX \citep{Gustenetal2006} during several runs from 2008 until 2013. The \champp\ observations targeted the \citwoone\ line at $809\,$GHz with the high-frequency array, with the $^{12}$CO $6-5$ line observed simultaneously in the low frequency array. The FLASH observations targeted \cionezero. Deep FLASH follow up integrations on the \cionezero\ line towards HD~100546, HD~97048 and TW~Hya were carried out in 2014. The backends used were AFFTS (\champp, with a highest resolution channel spacing of $0.18$~MHz or $0.11$~km$\,$s$^{-1}$ at $492$~GHz) and XFFTS (FLASH, $0.04$~MHz or $0.02$~\kms). The targeted \ci\ and CO lines are summarized in Table~\ref{tab:lines}. Diffraction-limited beam sizes range from $8''$ to $13''$. Most \champp\ observations were taken in single pointing mode, with a typical wobbler switch of $2'$ in azimuth. \champp\ has a central pixel, with the other six radially offset by $18''$ ($14''$) at $660$~GHz ($850$~GHz) in a hexagonal pattern. Raster mapping was used for AA~Tau, BP~Tau, DL~Tau, GG~Tau, RY~Tau, HD~36112 and Haro~65~B. The additional sky coverage was used to check for extended emission.

Typical survey observations had $10$ to $400$ scans with individual exposure times of $0.1$ to $0.5$~minutes and on-source exposure times of $10$ to $60$~minutes. The column of precipitable water vapour was typically $0.3$~mm~\htwoo, but as low as $0.1$ and as high as $0.7$ during some observations. Smoothed to $dv=0.17\,$\kms, the median RMS noise was $0.31\,$K for CO~$6$--$5$, $0.18\,$K for \cionezero\ and $0.67\,$K for \citwoone. Initial processing was done using the APECS software \citep{Mudersetal2006}. Baseline subtraction and other higher level reductions were done with GILDAS/CLASS\footnote{\texttt{http://www.iram.fr/IRAMFR/GILDAS}}. Telescope parameters were obtained from \citet{Gustenetal2006}. The intensities, \tastar, are corrected for atmospheric and radiative loss and spillover; they can be converted to main beam temperature via \tmb$\rm=(F_{eff}/\eta_{mb})\times$\tastar, where $\rm F_{eff}=0.95$. Based on \citet{Gustenetal2006}, the Kelvin-to-Jansky conversion and main beam efficiency at 650~GHz are $57$~Jy$\,$K$^{-1}$ and $0.53$, respectively. At 812~GHz, they are $70$~Jy$\,$K$^{-1}$ and $0.43$. At $491$~GHz, the conversion factor is $49$~Jy$\,$K$^{-1}$ and $\eta_{mb}=0.59$.

All sources and flux values from our survey are listed in Table~\ref{tab:sources} and shown in Fig.~\ref{fig:mainplot3}, including the detections of \cionezero\ towards TW~Hya and HD~100546. The spectra are presented in Fig.~\ref{fig:allspec}, and an example of extended emission in Fig.~\ref{fig:dmtau}.

\subsection{Complementary data}

To complement our \cionezero\ data for HD~142527, the \cosixfive\ and \citwoone\ lines from \citet{Casassusetal2013} are included in Table~\ref{tab:sources}. We also use observations of the \cii~$J=3/2$--$1/2$ transition at $158\,\mu$m towards a number of disks, obtained with the \emph{Herschel}/PACS low-resolution spectrometer by the GASPS \citep{Thietal2010, Meeusetal2012, Dentetal2013, Howardetal2013} and DIGIT \citep{Fedeleetal2013a} key programmes.

\begin{table*}[!ht]
\tiny
\centering
\caption{Summary of the disks and line fluxes observed with APEX.}
\label{tab:sources}
\begin{tabular}{l c c c | c c | c c | c c }
\hline\hline
Source	&	Spectral	& $d$	& $i_{\rm disk}$	&	\multicolumn{2}{c}{\cosixfive}		&	\multicolumn{2}{c}{\cionezero}		&	\multicolumn{2}{c}{\citwoone}		\\
		& 	type$^{\star}$		& 		& 			&	Flux$\,\pm\,3\sigma$	& RMS		&	Flux$\,\pm\,3\sigma$	& RMS		&	Flux$\,\pm\,3\sigma$	& RMS	 	\\
		& 					& (pc)	& ($^{\circ}$)	&	(K$\,$\kms)	& (K)			&	(K$\,$\kms)	& (K)			&	(K$\,$\kms)	& (K)	 	\\
\hline
\object{AA Tau}		& K7$^{\rm s1}$		& $140$					& $71^{\rm i1}$	&	$\leq 0.61$		& $0.58$	& $0.2\pm0.2$(e?)	& $0.18$		& $\leq 1.5$	& $1.4$	\\
\object{AS 205}		& K5$+$K7$+$M0$^{\rm s12}$		& $125^{\rm d1}$		& $15+20^{\rm i2}$	&	$2.8\pm0.3$		& $0.33$	&	\noobs	&\noobs	&	$\leq 0.63$	& $0.66$		\\
\object{AS 209}		& K4$^{\rm s2}$	& $119\pm6^{\rm d2}$		& $56^{\rm i3}$	&	$\leq 0.3$		& $0.26$	&	\noobs	&\noobs	&	$\leq 0.36$	& $0.38$	\\
\object{BP Tau}		& K7$^{\rm s3}$		& $140$					& $30^{\rm i4}$	&	$\leq 0.4$		& $0.42$	&	\noobs	&\noobs	&	$\leq 1.1$	& $1.1$		\\
\object{CS Cha}	& K6$^{\rm s4}$		& $150^{\rm d3}$			& $60^{\rm i5}$	&	$\leq 0.3$		& $0.34$	&	\noobs	&\noobs	&	$\leq 0.71$	& $0.73$	\\
\object{CV Cha}	& K0$^{\rm s5}$		& $150^{\rm d3}$			& $35^{\rm i6}$	&	$\leq 0.3$		& $0.33$	&	\noobs	&\noobs	&	$\leq 0.64$	& $0.66$	\\

\object{DL Tau}		& K7				& $140$					& $38^{\rm i7}$	&	$\leq 0.4$		& $0.36$	&	\noobs	&\noobs	&	$\leq 1.0$	& $1.0$		\\
\object{DM Tau}	& M1$^{\rm s6}$		& $140$					& $32^{\rm i4}$	&	$0.5\pm0.2$		& $0.22$	&	$0.15\pm0.14$(e?)	& $0.09$	&	$\leq 0.54$	& $0.51$		\\
\object{GG Tau A}	& M0$+$M2$+$M3$^{\rm s11}$& $140$			& $37^{\rm i8}$	&	$0.7\pm0.4$		& $0.41$	&	\noobs	&\noobs	&	$\leq 1.0$	& $1.0$		\\
\object{GM Aur} 	& K3$^{\rm s6}$		& $140$					& $56^{\rm i4}$	&	$\leq 0.5$		& $0.52$	&	\noobs	&\noobs	&	$\leq 0.91$	& $0.94$		\\

\object{HD 100453}	& A9				& $122\pm10^{\rm d4}$	& \noobs				&	$\leq 0.2$		& $0.22$	&	\noobs	&\noobs	&	$\leq 0.4$	& $0.38$		\\
\object{HD 100546}	& B9				& $97\pm4^{\rm d4}$		& $44^{\rm i9}$	&	$6.4\pm0.3$	& $0.18$	&	$0.69\pm0.15$	& $0.12$	&	$\leq 0.86$ & $0.54$		\\
\object{HD 104237}	& A4				& $116\pm5^{\rm d4}$		& $18^{\rm i10}$	&			\noobs		&	\noobs	&	$\leq 0.18$	& $0.02$	&	\noobs	& \noobs		\\
\object{HD 139614}	& A8				& $140\pm5^{\rm d6}$		& $20^{\rm i11}$	&	$\leq 0.3$		& $0.29$	&	\noobs	&\noobs	& $\leq 0.61$	& $0.63$		\\

\object{HD 141569}	& A0				& $116\pm7^{\rm d4}$		& $51^{\rm i12}$	&	$\leq 0.2$		& $0.16$	&	\noobs	&	\noobs& $\leq 0.6$	& $0.63$		\\
\object{HD 142527}	& F6				& $230\pm50^{\rm d4}$	& $20^{\rm i13}$	&	$1.5\pm0.9^{\rm C13}$	& 	\noobs	&	$\leq 0.23$	& $0.21$	&	$\leq 0.6^{\rm C13}$	& 	\noobs	\\
\object{HD 144668}	& A7				& $160\pm15^{\rm d4}$	& $58^{\rm i14}$	&	$2.2\pm0.1$(e)	& $0.08$	&	\noobs	&	\noobs	&	$\leq 0.53$	& $0.55$		\\
\object{HD 163296}	& A1				& $120\pm10^{\rm d4}$	& $45^{\rm i15}$	&	\noobs				&	\noobs	&	$\leq 0.4$	& $0.39$	&	\noobs	&	\noobs	\\
\object{HD 169142}	& A5				& $145\pm5^{\rm d6}$		& $8^{\rm i10}$	&	$1.1\pm0.2$		& $0.21$	&	$\leq 0.3$	& $0.27$	&	$\leq 0.4$	& $0.37$		\\

\object{HD 36112} (MWC 758)	& A8				& $280\pm55^{\rm d4}$	& $21^{\rm i16}$	&	$\leq 0.3$		& $0.32$	&	\noobs	&	\noobs&	$\leq 0.94$	& $0.97$		\\
\object{HD 38120}	& A5$^{\rm s13}$			& $480\pm175^{\rm d4}$	& \noobs	&	$\leq 0.2$		& $0.19$	&	\noobs	&	\noobs&	$\leq 0.48$	& $0.50$		\\
\object{HD 50138}	& B9?				& $390\pm70^{\rm d4}$	& \noobs	&	$\leq 0.3$		& $0.29$	&	\noobs	&	\noobs &	$\leq 0.43$	& $0.45$		\\
\object{HD 97048}	& A0e 				& $160\pm15^{\rm d4}$	& $43^{\rm i17}$	&	$7.0\pm0.3$		& $0.30$	&	$1.4\pm0.3$	& $0.07$	&	$2.1\pm0.9$	& $0.75$		\\

\object{Haro 6-5 B}	& M4				& $145^{\rm d7,d8}$		& $74^{\rm i18}$	&	$\leq 0.5$		& $0.49$	&	\noobs	&	\noobs&	$\leq 1.3$	& $1.4$		\\
\object{IM Lup}		& M0				& $155\pm8^{\rm d2}$		& $54^{\rm i19}$	&	$0.6\pm0.2$		& $0.23$	&	$\leq 0.2$	& $0.19$	&	$\leq 0.44$	& $0.46$		\\
\object{LkH$\alpha$ 264}	& K5.5$^{\rm s7}$	& $360\pm30^{\rm d9}$& $20^{\rm i20}$	&	$\leq 0.2$		& $0.15$	&	$\leq 0.4$	& $0.40$	&	$\leq 0.4$	& $0.38$		\\
\object{RY Tau}		& K1$^{\rm s3}$		& $140$					& $65^{\rm i21}$	&	$\leq 0.3$		& $0.31$	&	\noobs	&\noobs	&	$\leq 1.1$	& $1.1$		\\
\object{Sz 33}		& M3.5$+$M7$^{\rm s8}$	& $150^{\rm d3}$		& \noobs	&	$1.4\pm0.4$		& $0.43$	&	\noobs	&	\noobs &	$\leq 0.8$	& $0.67$		\\
\object{Sz 41}		& K7$+$M2.5$^{\rm s8}$	& $150^{\rm d3}$		& \noobs	&	$\leq 0.5$		& $0.43$	&	\noobs	& \noobs	&	$\leq 0.8$	& $0.86$		\\

\object{T~Cha}		& K0$^{\rm s2}$		& $108\pm9^{\rm d10}$	& $67^{\rm i22}$	&	$\leq 0.2$		& $0.23$	&	\noobs	&	\noobs&	$\leq 0.8$	& $0.78$		\\
\object{TW Hya}	& K7$^{\rm s9}$	& $56\pm7^{\rm d11}$		& $7^{\rm i23}$	&	$1.0\pm0.3$		& $0.28$	&	$0.05\pm0.03$	& $0.10$	&	$\leq 0.49$	& $0.51$	\\
\object{V806 Tau} 	& M0				& $140$					& \noobs	&	$3.1\pm0.8$		& $0.83$	&	\noobs	& \noobs	&	$\leq 3.0$	& $3.1$		\\
\object{V892 Tau}	& A0				& $140$					& $60^{\rm i24}$	&	$4.6\pm0.8$(e)	& $0.77$	&	\noobs	&	\noobs &	$\leq 2.4$	& $2.5$	 	\\
\object{VZ Cha} 	& K7$^{\rm s5}$		& $150^{\rm d3}$			& \noobs	&	$\leq 1.1$		& $1.0$	&	\noobs	&	\noobs &	$\leq 2.6$	& $2.3$		\\
Oph IRS 48 (\object{WLY 2-48})	& A0$^{\rm s10}$		& $139\pm6^{\rm d12}$	& $50^{\rm i25}$	&		\noobs		& \noobs	&	$0.8\pm0.4$(e?)	& $0.18$	&	\noobs	&	\noobs	\\
\hline
Median		&		&	$140$	&	$45$	&		&	$0.31$	&		&	$0.21$	&		&	$0.66$	 \\
\hline
\end{tabular}

\flushleft
\emph{Notes: }All uncertainties and upper limits are at $3\sigma$ confidence. Extended emission is indicated with `(e?)'. The intensity scale is \tastar\ (K). Flux upper limits were calculated as RMS${\times}\sqrt{N_{\rm chan}{\times}\delta v}$ over a linewidth of $\delta v=5$~\kms, and $3\sigma$ RMS values are given for a $dv=0.17$~\kms\ channel spacing. Spectral types without references are from SIMBAD. $^{\star}$ -- components of arcsecond-scale multiples are separated with a plus sign; $^{\rm C13}$ -- \citet{Casassusetal2013}; $^{\rm T15}$ -- \citet{Tsukagoshietal2015}; $^{\rm d1}$ -- \citet{Pontoppidanetal2011}; $^{\rm d2}$ -- \citet{Lombardietal2008}; $^{\rm d3}$ -- \citet{KnudeHog1998}; $^{\rm d4}$ -- \citet{vanLeeuwen2007}; $^{\rm d5}$ -- \citet{vanBoekeletal2005}; $^{\rm d6}$ -- \citet{AckevandenAncker2004}; $^{\rm d7}$ -- \citet{Torresetal2009}; $^{\rm d8}$ -- \citet{Torresetal2012}; $^{\rm d9}$ -- \citet{Anderssonetal2002}; $^{\rm d10}$ -- \citet{Torresetal2008}; $^{\rm d11}$ -- \citet{Wichmannetal1998b}; $^{\rm d12}$ -- \citet{Mamajek2008}; $^{\rm i1}$ -- \citet{Coxetal2013}; $^{\rm i2}$ -- \citet{Salyketal2014}; $^{\rm i3}$ -- \citet{KoernerSargent1995b}; $^{\rm i4}$ -- \citet{Simonetal2000}; $^{\rm i5}$ -- \citet{Espaillatetal2007}; $^{\rm i6}$ -- \citet{Hussainetal2009}; $^{\rm i7}$ -- \citet{Guilloteauetal2011}; $^{\rm i8}$ -- \citet{Guilloteauetal1999}; $^{\rm i9}$ -- \citet{Walshetal2014}; $^{\rm i10}$ -- \citet{Malfaitetal1998b}; $^{\rm i11}$ -- \citet{Matteretal2014}; $^{\rm i12}$ -- \citet{Weinbergeretal1999}; $^{\rm i13}$ -- \citet{Canovasetal2013}; $^{\rm i14}$ -- \citet{Preibischetal2006}; $^{\rm i15}$ -- \citet{deGregorioMonsalvoetal2013}; $^{\rm i16}$ -- \citet{Isellaetal2010}; $^{\rm i17}$ -- \citet{Doucetetal2007}; $^{\rm i18}$ -- \citet{Starketal2006}; $^{\rm i19}$ -- \citet{Panicetal2009a}; $^{\rm i20}$ -- \citet{Carmonaetal2007b}; $^{\rm i21}$ -- \citet{McClearyetal2007}; $^{\rm i22}$ -- \citet{Huelamoetal2015}; $^{\rm i23}$ -- \citet{Qietal2004}; $^{\rm i24}$ -- \citet{Monnieretal2008}; $^{\rm i25}$ -- \citet{vanderMareletal2013}; $^{\rm s1}$ -- \citet{WhiteGhez2001}; $^{\rm s2}$ -- \citet{Torresetal2006}; $^{\rm s3}$ -- \citet{Bertoutetal2007}; $^{\rm s4}$ -- \citet{Luhman2004}; $^{\rm s5}$ -- \citet{Torresetal2006}; $^{\rm s6}$ -- \citet{Bertoutetal2007}; $^{\rm s7}$ -- \citet{Luhman2001}; $^{\rm s8}$ -- \citet{Daemgenetal2013}; $^{\rm s9}$ -- \citet{RucinskiKrautter1983}; $^{\rm s10}$ -- \citet{Brownetal2012}; $^{\rm s11}$ -- \citet{DiFolcoetal2014}; $^{\rm s12}$ -- \citet{Eisneretal2005}; $^{\rm s13}$ -- \citet{Meeusetal2012}
\end{table*}

\section{Observational results}\label{sec:obsresults}

The APEX spectra of $37$ disks are shown in Fig.~\ref{fig:allspec} and the measurements are summarized in Table~\ref{tab:sources}. Not all lines were observed towards all disks. Emission in the \cosixfive\ line was detected (observed) towards $13$ ($33$) sources; the corresponding numbers are $6$ ($12$) for the \cionezero\ line and $1$ ($33$) for the \citwoone\ line. When considering detections and upper limits, sources of all spectral types cover a similar range in line flux, although the CO and \ci\ detections towards Herbig disks are typically a factor of a few stronger than towards T~Tauri disks.

The \cosixfive\ lines are single-peaked and narrow, except for three sources which show wide, double-peaked lines. The first is HD~100546, where the line displays the same blue-over-red peak asymmetry seen in lower-$J$ CO observations \citep[e.g.,][where the CO~$6$--$5$ line was also previously shown]{Panicetal2010}. The second is HD~97048, which has a filled-in line center suggestive of a non-disk emission component. The third is IM~Lup, a very large T~Tauri disk \citep{vanKempenetal2007, Panicetal2009a}. The \cosixfive\ line has been observed previously towards a number of disks. The upper limit obtained by \citet{vanZadelhoffetal2001} towards TW~Hya lies $30$\% above the detection reported in Table~\ref{tab:sources}. A comparison of the six overlapping sources with \citet{Thietal2001} shows good consistency, with APEX yielding improved upper limits and detections. In the case of GM~Aur, our upper limit lies a factor of $1.5$ below the detection from 2001, but is within its errorbars. The \cosixfive\ flux detected towards HD~142527 by \citet{Casassusetal2013}, $1.5\pm0.9$~K$\,$\kms, is roughly in the middle of the range of values across our entire sample. The values reported for HD~100546 and HD~97048 in the much larger \emph{Herschel}/SPIRE beam by \citet{vanderWieletal2014} are also consistent with our detections.

In most sources where \ci\ emission is detected, the emission is extended or contaminated by the reference position. The exceptions are HD~100546 and TW~Hya. These two are the first two unambiguous detections of \ci\ in protoplanetary disks, with a \tastar\ line flux of $0.49$~K$\,$\kms\ (S/N${=}10$) towards HD~100546 and $0.05$~K$\,$\kms\ (S/N${=}3$) towards TW~Hya. Gaussian fit parameters for both detections are given in Table~\ref{tab:cidetections}. We also detect a strongly asymmetric double-peaked line towards HD~97048. This detection likely has a disk contribution, but it is difficult to quantify because of contamination issues affecting the line center. The \cionezero\ line is detected towards Oph~IRS~48 (also known as WLY~2-48), but the line profile is asymmetric, with emission redshifted from the rest velocity of $4.6\,$km$\,$s$^{-1}$ and absorption on the red side. The disk contribution can thus not be determined at the moment.

\begin{table}[!ht]
\centering
\caption{Gaussian fit parameters for the definitive \cionezero\ detections.}
\label{tab:cidetections}
\begin{tabular}{ c c c c c }
\hline\hline
Source	& Flux	& Peak	& $v_{\rm lsr}$	& Width\\
		& (K$\,$km$\,$s$^{-1}$)	&	(K)	&	(km$\,$s$^{-1}$)	& (km$\,$s$^{-1}$)	\\
\hline
HD~100546	& $0.69\pm0.15$	&	$0.15\pm0.12$	&	$5.6\pm0.6$	&	$4.4\pm0.9$	\\
TW~Hya		& $0.05\pm0.03$	&	$0.11\pm0.10$	&	$2.7\pm0.3$	&	$0.41\pm0.36$	\\	
\hline
\end{tabular}
\flushleft
\emph{Notes: }All uncertainties are $3\,\sigma$.
\end{table}

The \ci\ transitions have recently been surveyed with \emph{Herschel}/SPIRE by \citet{vanderWieletal2014}, who report no firm detections. Generally, the \cionezero\ limits from APEX have better sensitivity than SPIRE, while SPIRE provides deeper limits to the $2$--$1$ line. However, SPIRE could not spectrally or spatially resolve disk emission. Its beam at $809\,$GHz was $26''$, compared to $8''$ for APEX. The tentative SPIRE detection of \citwoone\ for HD~100546 lies within errorbars of our upper limit, which is a re-evaluation of the data reported by \citet{Panicetal2010}. For HD~100453, HD~169142, HD~36112 and HD~50138, the $2$--$1$ limits are similar. We also detect \citwoone\ towards HD~97048, but above the upper limit of \citet{vanderWieletal2014}, suggesting that the emission is extended and was subtracted out in the SPIRE analysis where off-source spaxels were used as a reference pointing. This also makes our \cionezero\ detection towards that source suspect. For RY~Tau, the SPIRE limit is a factor of five deeper than the APEX one. The proximity of TW~Hya allowed our deep APEX integrations to yield a detection of the faint \cionezero\ transition and the lowest distance-corrected upper limit on the $2$--$1$ line in our survey.

\begin{figure}[!ht]
\includegraphics[clip=, width=1.0\columnwidth]{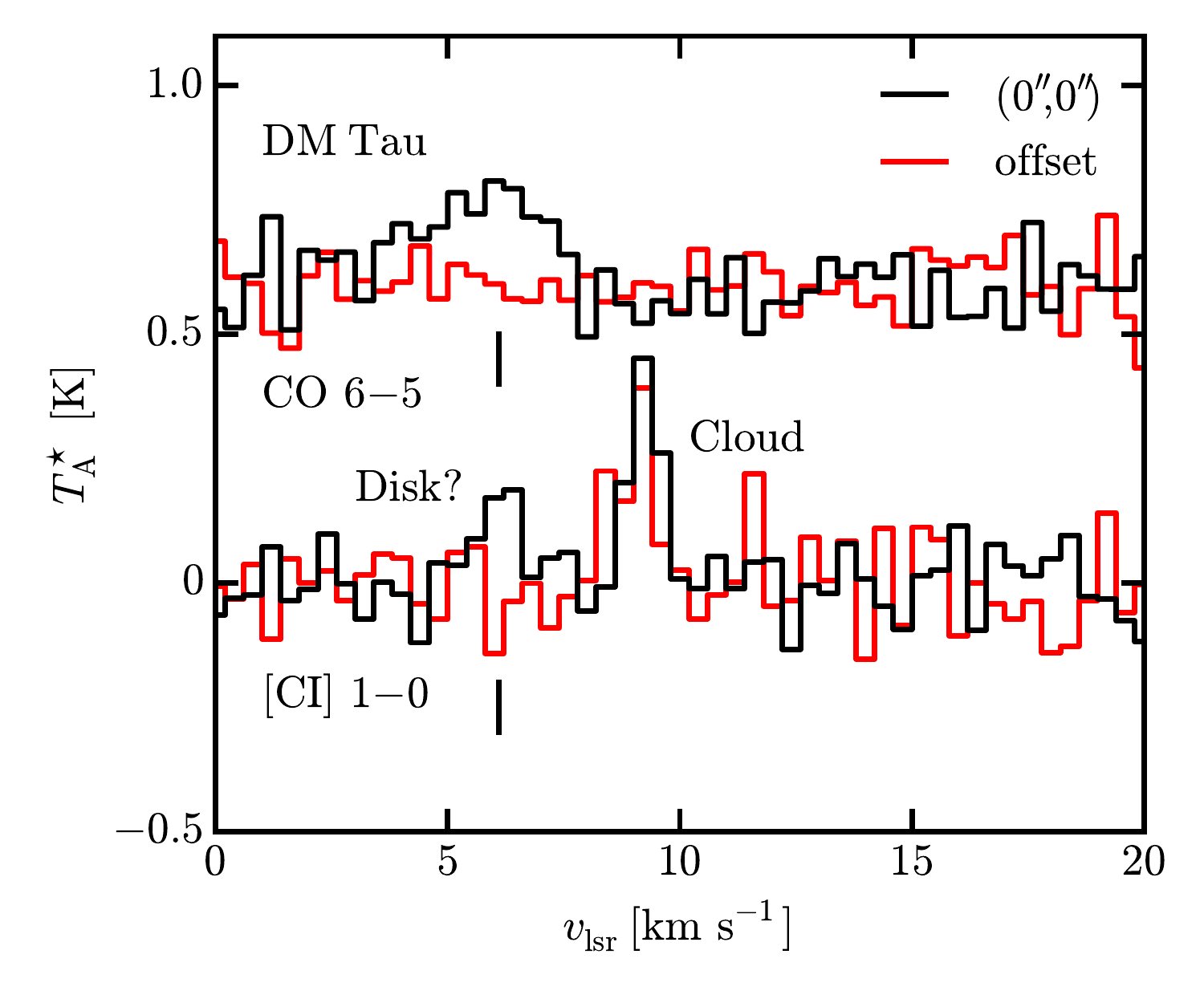}
\caption{The \cionezero\ and \cosixfive\ lines observed towards DM~Tau (black) and typical offset positions (red; $+30'',0''$ for \ci, $+18'',+6''$ for CO). The \ci\ line towards the source is single-peaked and narrower than \cosixfive. The spectra are binned to $0.4\,$km$\,$s$^{-1}$ per channel.}
\label{fig:dmtau}
\end{figure}

DM~Tau shows purely on-source \cosixfive\ emission and extended \ci\ emission, as shown in Fig.~\ref{fig:dmtau}. For CO, a broad line consistent with the disk inclination is detected on-source while the off-source position is clean. The on-source \ci\ emission has two narrow components, one strong and narrow peak due to extended emission at $(9.3\pm0.1)\,$km$\,$s$^{-1}$, also seen in the reference spectrum, and the other at $(6.3\pm0.2)\,$km$\,$s$^{-1}$ probably originating in a compact envelope. The latter matches the CO detection, but is narrower. Our on-source detection of \cionezero\ is within the errorbars of the value from \citet{Tsukagoshietal2015}, however the \ci\ line from this inclined disk is single-peaked and  narrower than the CO line, suggesting an envelope or wind contribution. The same narrow kinematic components are seen for this source in the low-$J$ CO lines \citep{GuilloteauDutrey1994}. Aside from having the $9\,$km$\,$s$^{-1}$ component, the low-$J$ CO isotopolog lines at $6\,$km$\,$s$^{-1}$ are double-peaked and the $^{12}$CO$\,2$--$1$ line shows an additional narrow emission peak at the systemic velocity. The \cosixfive\ line appears to originate purely in the disk. Follow-up observations are needed to firmly establish the origin of the \cionezero\ emission towards DM~Tau.

A single-peaked \cionezero\ line is also marginally detected towards AA~Tau. No offset position was observed, but given that this disk is seen edge-on and accordingly has very broad emission lines \citep[e.g.][]{Brownetal2013}, the narrow line is likely cloud emission.

The \cosixfive\ line towards HD~144668 is uniform across all seven \champp\ pixels, with a flux increase towards northern positions, suggesting the line originates entirely in an extended cloud. The T~Tauri systems AS~205, Sz~33 and V806~Tau have very strong \cosixfive\ emission, but display no emission in the off-source \champp\ pixels, ruling out extended emission. Their single-peaked on-source line profiles suggest a contribution from a remnant envelope or a disk wind \citep{Pontoppidanetal2011, Salyketal2014}. AS~205 is a hierarchical T~Tauri triple within $1.3''$ \citep{Eisneretal2005}. For AS~205A, single-peaked profiles are also seen in ro-vibrational CO lines \citep{Bastetal2011, Brownetal2013}.

\begin{table*}[!ht]
\centering
\caption{The atomic and molecular lines used in this paper.}
\label{tab:lines}
\begin{tabular}{ c c c c c c c c }
\hline\hline
Species	& Transition	& $E_{\rm u}$	& $A_{\rm ul}$	& $n_{\rm crit}$ (\htwo\ at $100$~K)	& $\nu$	& $\theta_{\rm beam}$	& Instrument	\\
		&			& (K)		& ($^{-1}$)		& (cm$^{-3}$)		& (GHz)	& ($''$)				&	\\
\hline
\catom		& \ci~$^{3}$P$_{1}$--$^{3}$P$_{0}$ & $23.6$	& $7.880\times 10^{-8}$	& $5\times 10^{2}$	& $492.16065$	& $13$	& APEX/FLASH	\\

\catom		& \ci~$^{3}$P$_{2}$--$^{3}$P$_{1}$ & $62.5$	& $2.650\times 10^{-7}$	& $5\times 10^{2}$	& $809.34197$	& $8$	& APEX/\champp \\
CO		& CO~$6$--$5$	& $116.2$	& $2.137\times10^{-5}$	& $3\times 10^{5}$	& $691.47308$	& $10$	& APEX/\champp \\

\cplus	&	\cii~$^{2}$P$_{3/2}$--$^{2}$P$_{1/2}$	&	$91.21$	&	$2.300\times10^{-6}$		&	$5\times10^{3}$	&	$1900.5369$		&	$11$	& \herschel/PACS \\

\hline
\end{tabular}
\end{table*}

\section{Modelling}\label{sec:modelling}

Our next goals are to relate the observed \ci\ and CO emission to the elemental \chgas\ ratio in the warm atmosphere of a disk, where surface chemistry is unimportant, and to understand the low detection rates and upper limits in our survey. To this end, we make use of the \texttt{DALI} physical-chemical code \citep{Brudereretal2012, Brudereretal2013}. Starting with a parameterized gas and dust density distribution (Section~\ref{sec:diskparams}), and an input stellar spectrum (Section~\ref{sec:stellarparams}), \texttt{DALI} first solves the continuum radiative transfer to determine the UV radiation field and the dust temperature. This yields an initial guess for the gas temperature, which is the starting point of an iterative sequence in which the chemistry and thermal balance are calculated. Once the solution is converged, the code can output emission maps, spectra and integrated fluxes. We present below a grid of generic models, with parameter ranges covering the source sample. All observables are hereafter normalized to $140\,$pc. Tailored models for TW~Hya and HD~100546, with detailed fitting of the carbon abundance, will be presented in a companion paper (Kama et al. submitted).

\subsection{Disk parameters}\label{sec:diskparams}

The disk density structure in our version of \texttt{DALI} is fully parameterized. The gas-to-dust mass ratio is \gdrat. The surface density has the standard form of a power law with an outer exponential taper:
\begin{equation}
\Sigma_{\rm gas} = \Sigma_{\rm c}\cdot \left( \frac{r}{R_{\rm c}} \right)^{-\gamma} \cdot \exp{\left[ - \left(\frac{r}{R_{\rm c}}\right)^{2-\gamma} \right]}.
\end{equation}

To simulate an inner cavity, material can be removed inside of some radius $r_{\rm hole}$. The scaleheight angle, $h$, at distance $r$ is given by $h(r)=h_{\rm c}\,(r/R_{\rm c})^{\psi}$, such that the scaleheight is $H=h\cdot r$, and the vertical density structure of the small grains is
\begin{equation}\label{eq:rhosmall}
\rho_{\rm d,small} = \frac{(1-f)\,\Sigma_{\rm dust}}{\sqrt{2\,\pi}\,r\,h} \times \exp{ \left[ -\frac{1}{2} \left( \frac{\pi/2 - \theta}{h} \right)^{2} \right] },
\end{equation}

where $f$ is the mass fraction of large grains and $\theta$ is the opening angle from the midplane as viewed from the central star. The settling of large grains is prescribed as a fraction $\chi \in{(0,1]}$ of the scaleheight of the small grains, so the mass density of large grains is similar to Eq.~\ref{eq:rhosmall}, with $f$ replacing $(1-f)$ and $\chi\,h$ replacing $h$. The vertical distribution of gas is calculated in each grid cell as $\rho_{\rm gas}=\Delta_{\rm g/d}\times \rho_{\rm d,small}\times [1+f/(1-f)$]. The latter factor adds the mass of large grains as if they were not settled, to preserve the global \gdrat. The ranges and fiducial values of all parameters in our model grid are given in Table~\ref{tab:gridparams}. The range of $R_{\rm c}$, $\psi$ and $\gamma$ are guided by results from homogeneous sample fitting studies \citep{Andrewsetal2009, Andrewsetal2010}. The main fiducial disk model has a total mass of $10^{-2}\,$\msol\ and a reference surface density of $\Sigma_{\rm c}=5.5\,$g$\,$cm$^{-2}$ at $R_{\rm c}=50\,$au. We also calculate some models for a very small fiducial disk, with $R_{\rm c}=10\,$au and $\Sigma_{\rm c}=140\,$g$\,$cm$^{-2}$.

\begin{table}[!ht]
\centering
\caption{The fiducial values and grid ranges of the model parameters.}
\label{tab:gridparams}
\begin{tabular}{ c c c c }
\hline\hline
Parameter				&	Fiducial	&	Range	& Units		\\
\hline
$\gamma$			&	$1.0$	&	[$\,0.8$, $1.5\,$]		&	\\
$R_{\rm c}$			&	$50$, $10$	&	[$\,10$, $130\,$	] &	(au)	\\ 
$\Sigma_{\rm c}$		&	$5.5$	&	[$\,0.055$, $55\,$]				&	(g$\,$cm$^{-2}$)	\\
$\Sigma_{\rm c}(R_{\rm c}{=}{\rm 10\,}$au)		&	$140$	&	single value				&	(g$\,$cm$^{-2}$)	\\
$M_{\rm disk}$			&	$10^{-2}$	&	[$\,10^{-4}$, $10^{-1}\,$]	&	(\msol)\\
\gdrat				&	$100$	&	[$\,10$, $1000\,$]	&	\\
$h_{\rm c}$							&	$10$		&	[$\,0.05$, $0.20\,$]		&	\\
$\psi$								&	$0.2$	&	[$\,0.05$, $0.3\,$]		&	\\
$\chi$ 						&	$0.5$	&	[$\,0.1$, $1.0\,$]	&	\\
$f$							&	$0.90$	&	[$\,0.20$, $0.99\,$]		&	\\
$r_{\rm hole}$							&	$0$		&	[$\,0$, $75\,$]	&	(au)	\\
$L_{\rm X}$, $T_{X}$					&	$7.94\times 10^{28}$	& single value	&	(erg$\,$s$^{-1}$)\\
$L_{\rm X}$, $T_{X}$					&	$7\times10^{7}$	& single value	&	(K)	\\
$\zeta_{\rm cr}$						&	$5$		&	[$\,0.05$, $500\,$]	&	($10^{-17}\,$s$^{-1}$	)\\
\hline
\chgas($R_{\rm c}{=}{\rm 50, 10}\,$au)								&	$1.35$	&	[$\,0.01$, $2\,$]	& ($\times 10^{-4}$)	\\
\hline
$i$									&	$60$		&	[$\,0$, $90\,$]	&	($^{\circ}$)	\\
\hline
\end{tabular}
\flushleft
\emph{Notes: }Unless explicitly noted, all variations are around the $R_{\rm c}{=}{\rm 50}\,$au fiducial model. For variations of $R_{\rm c}$, we keep the disk mass fixed at $10^{-2}\,$\msol\ and accordingly adjust $\Sigma_{c}(R_{\rm c})$.
\end{table}

\subsection{Stellar parameters and accretion rates}\label{sec:stellarparams}

The stellar spectra were approximated as pure blackbodies. For consistency, the \teff\ of a star was uniquely related to a combination of mass, radius and luminosity using the PISA pre-main-sequence evolutionary tracks of \citet{Tognellietal2011} at a model age of $5\,$Myr. We adopt a \teff~$=10000\,$K star as representative for the Herbig~Ae/Be group (spectral types B, A, F) and a $4000\,$K star with UV-excess for the T~Tauri group (G, K, M). For \teff~$=4000\,$K, accretion at a rate of \mdot$\,=10^{-8}$~\msol$\,$yr$^{-1}$ was assumed to release energy in blackbody emission at the stellar photosphere at $T_{\rm acc}=10000$~K. The relevant stellar and ultraviolet luminosities are listed in Table~\ref{tab:uvlums}. Lyman~$\alpha$ radiation was not explicitly considered. This does not affect the photodissociation of CO or the -ionization of C$^{0}$, as only photons at $\lesssim 1100$~\AA\ are important. The fiducial X-ray luminosity adopted in the models, $L_{\rm X}=10^{29}$~erg$\,$s$^{-1}$, is close to the median value for the classical T~Tauri star sample from \citet{Neuhauseretal1995}.

\subsection{The chemical network}\label{sec:carbonchem}

The adopted chemical network is based on UMIST~06 \citep{Woodalletal2007}. It consists of $109$ species, including neutral and charged PAHs, and $1463$ individual reactions. In addition to two-body reactions, the code includes freezeout, thermal and photodesorption, and photodissociation and -ionization. Hydrogenation is the only grain surface reaction considered. This has no impact on the chemistry of the disk atmosphere species considered in this work.

We adopt a fiducial gas-phase elemental carbon and oxygen abundance of \chgas$\,=1.35\times10^{-4}$ and \ohgas$\,=2.88\times 10^{-4}$, respectively. These values are close to the medians observed in diffuse and translucent interstellar clouds by UV absorption lines of \cplus\ and O$^{0}$. The total elemental abundance of carbon is still subject to considerable uncertainties. The solar abundance is $2.69\times 10^{-4}$ \citep{Asplundetal2009}, whereas that of the material from which the Solar System formed is estimated at $2.88\times 10^{-4}$ \citep{Lodders2003}. Values as high as $4\times 10^{-4}$ have been advocated for the cosmic carbon abundance in the solar neighborhood \citep{Parvathietal2012}. \chgas\ sets the amount of carbon that is cycled in our model between the volatile phases, i.e., atoms, molecules and ices. The refractory carbonaceous dust reservoir is not explicitly considered. PAHs, which absorb stellar UV photons, are a major gas heating agent in the disk atmosphere \citep[e.g.,][]{BakesTielens1994, Habartetal2004}. Following typical values inferred from observations \citep[$0.1\ldots0.01$][and references therein]{Geersetal2006, Kamp2011}, the PAH abundance was set to $0.1$ of the interstellar value of [PAH]/[H]~$\sim10^{-7}$. When varying \chgas, we also vary \ohgas\ to keep the C/O ratio constant. Keeping \ohgas\ fixed would make the \catom\ abundance more strongly dependent on \chgas, as more oxygen would be available to bind up carbon.

\begin{table}[!ht]
\centering
\caption{The total and ultraviolet stellar luminosities.}
\label{tab:uvlums}
\begin{tabular}{ l c c c }
\hline\hline
Spectrum		& CO ph.dissoc.	& Broadband UV			&	Stellar	\\
			& $91.2\ldots 110$~nm	&	$91.2\ldots 200$~nm	&	$L_{\rm tot}$	\\
			&	(\lsol)		&	(\lsol)	&	(\lsol)	\\
\hline
4000~K								& $1.3\times 10^{-11}$	&	$5.8\times 10^{-6}$	&	$0.38$	\\
4000~K$+$UV							& $1.4\times 10^{-4}$	&	$1.2\times 10^{-2}$	&	$0.55$	\\
6000~K								& $5.7\times 10^{-6}$	&	$2.0\times 10^{-2}$	&	$9.88$	\\
6000~K$+$UV							& $1.6\times 10^{-4}$	&	$3.4\times 10^{-2}$	&	$10.1$	\\
8000~K								& $1.7\times 10^{-3}$	&	$5.7\times 10^{-1}$	&	$29.4$	\\
10000~K								& $2.6\times 10^{-2}$	&	$2.2\times 10^{\,0}$	&	$34.0$	\\
12000~K								& $2.3\times 10^{-1}$	&	$8.2\times 10^{\,0}$	&	$59.6$	\\
\hline
TW~Hya$^{r1}$						& $6.7\times 10^{-5}$	&	$7.5\times 10^{-3}$	&	${\approx}0.3$	\\
HD~100546$^{r2}$						& $1.5\times 10^{-2}$	&	$9.1\times 10^{\,0}$	&	${\approx}30$	\\
\hline
\end{tabular}
\flushleft
\emph{Notes: }Temperatures refer to the model stars described in Section~\ref{sec:stellarparams}, based on the pre-main sequence tracks of \citet{Tognellietal2011}. The excess UV for the $4000$ and $6000\,$K stars is for an accretion rate \mdot~$=10^{-8}\,$\msol$\,$yr$^{-1}$, with the energy released on the stellar surface at $10000\,$K effective temperature. The Lyman$\,\alpha$ luminosity of TW~Hya, integrated from $121$ to $122\,$nm, is $5.1\times 10^{-3}\,$\lsol. \emph{References: }$^{r1}$ -- \citet{Franceetal2014}, $^{r2}$ -- \citet{Brudereretal2012}.
\end{table}

\begin{figure*}[!ht]
\centering
\includegraphics[clip=, width=1.0\linewidth]{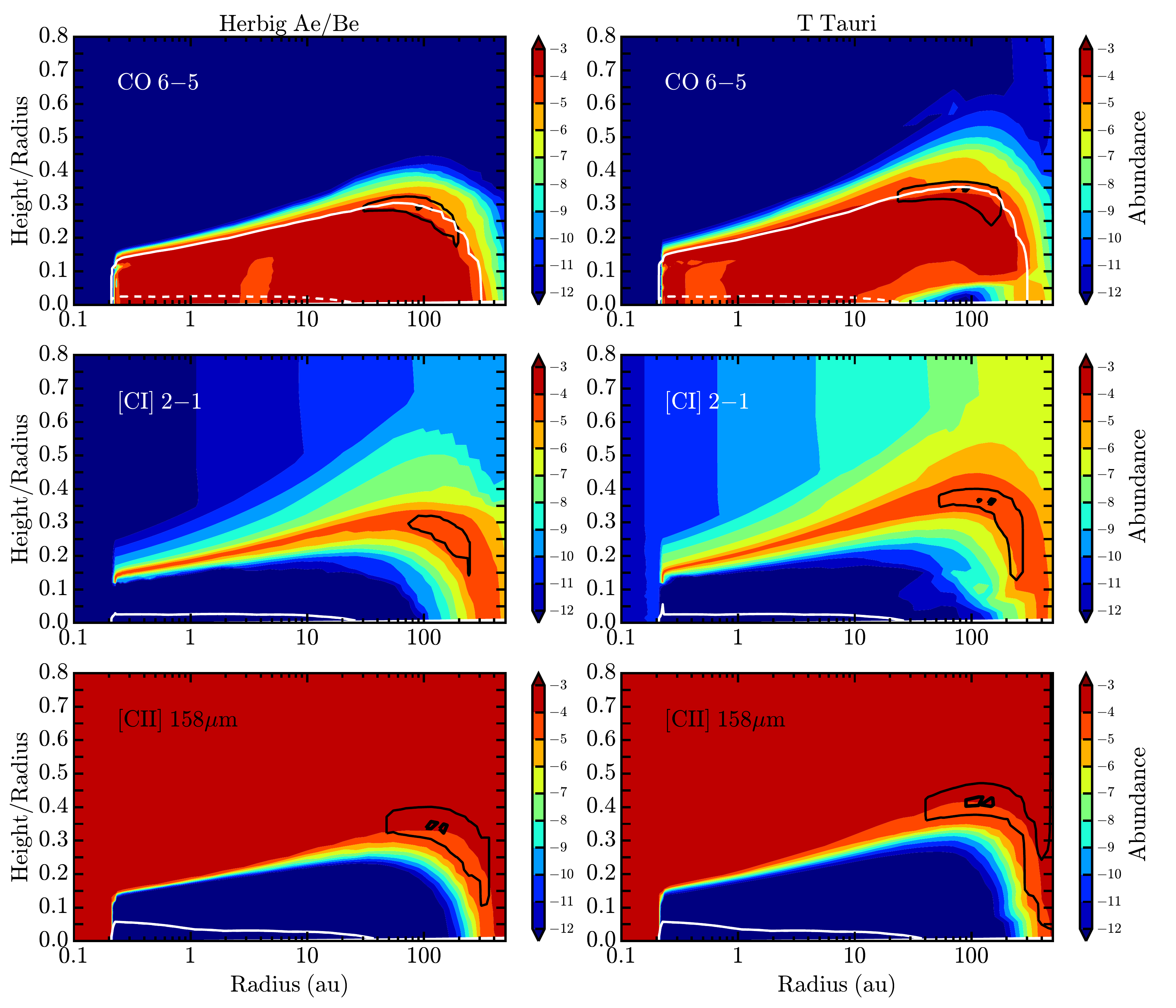}
\caption{Maps of the abundance (colour shading) and emission contribution (black lines) of the main gas-phase carbon reservoirs in a disk around a star with \teff\,$=10000$~K (left) and a \teff\,$=4000$~K star with $0.1$\,\lsol\ of UV excess (right). The solid black lines demarkate areas containing $75$\% (thinner line) and $25$\% (thicker) of the emission. White lines show the vertical $\tau=1$ surface for the line (solid) and continuum (dashed) emission.}
\label{fig:Xcbf}
\end{figure*}

\section{Modelling results}\label{sec:modresults} 

In Appendix~\ref{sec:modeldetails}, Figs.~\ref{fig:ngas} and \ref{fig:diskstruct}, we present the gas density, gas and dust temperature, and the extinction for the fiducial disk models. The fiducial Herbig disk is substantially warmer, notably the CO freezeout zone (\tdust$\,\lesssim25\,$K) is much larger in the T~Tauri disk than in the Herbig~Ae/Be one.

\subsection{Origin of the CO, \ci\ and \cii\ emission}

Abundance and emission contribution maps for the \cosixfive, \cionezero, \citwoone\ and \ciiline\ transitions are shown in Fig.~\ref{fig:Xcbf}. The exterior layers of a protoplanetary disk form a complex photodissociation region (PDR), but covering a much larger range of \ngas, $G_{0}$ and dust properties than standard PDRs in the interstellar medium \citep{TielensHollenbach1985, vanDishoecketal2006}. The general outcome is outside-in layering where the dominant gas-phase carbon reservoir switches from \cplus\ to \catom\ and then CO. The rotational lines of CO are optically thick to high $J$ levels and thus probe the surface area and temperature of the disk, while the optically thin \ci\ and \cii\ lines are ``carbon counters'' for the disk atmosphere. 

The emission of the \cosixfive\ transition ($E_{\rm u}=116.2$~K) originates primarily in the outer disk, but has contributions from the entire warm molecular layer. It is optically thick out to several hundred astronomical units. The \cionezero\ and \citwoone\ transitions have lower upper level energies compared to \cosixfive\ by a factor of a few, while the \catom\ atom has its peak abundance exterior to that of CO. Thus, the \ci\ lines predominantly originate in the surface layers of the outer disk, with a small contribution from the inner disk. Carbon is ionized everywhere exterior to the \catom\ layer and its optically thin emission probes the warm tenuous gas around the disk. Atomic carbon offers the clear advantage of counting carbon atoms in the warm atmosphere of the disk. The disk contribution can be verified with additional pointings and the line profile shape. \cii\ is more susceptible to nebular contamination and is usually spatially and spectrally unresolved, while CO usually suffers from optical depth issues. 

\begin{figure*}[!ht]
\centering
\includegraphics[clip=, width=0.97\linewidth]{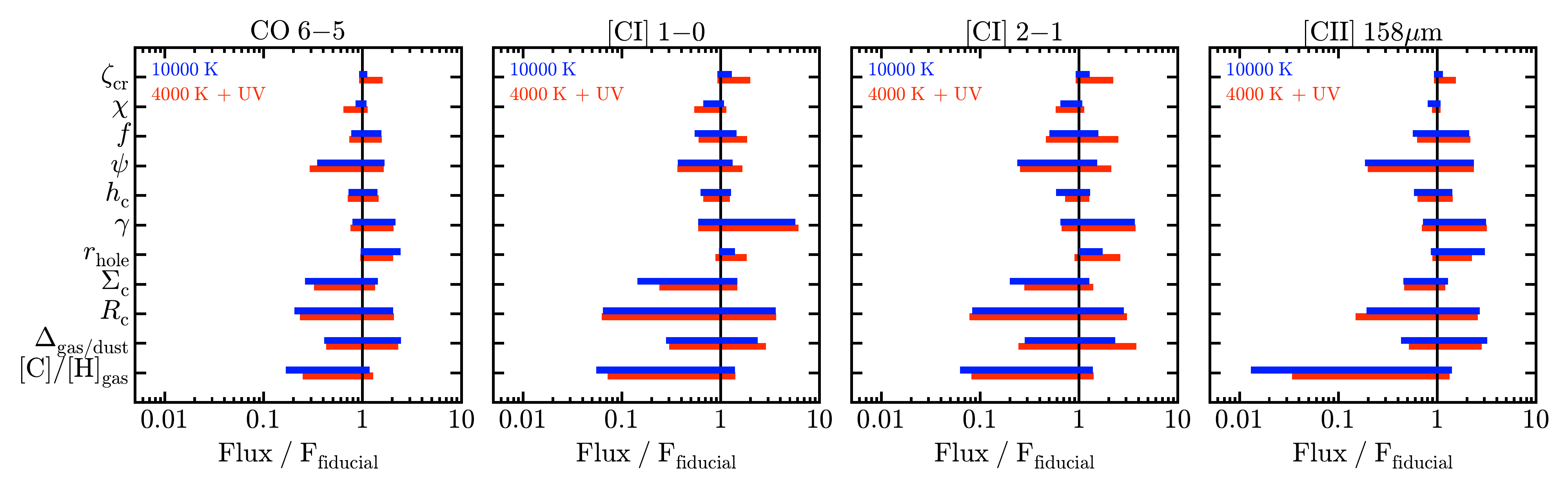}
\caption{The effect of model parameter variations on the line fluxes of carbon species in the disk atmosphere. Two central stars are considered: a $4000\,$K star with UV excess (red), and a pure $10000\,$K photosphere (blue). Horizontal bars show the range of variation resulting from the parameter values given in Table~\ref{tab:gridparams}. The fiducial model (black vertical line) is a disk with $R_{\rm c}=50$~au and with \chgas$\,=1.35\times 10^{-4}$, and the reference fluxes for the $4000\,$K models are, from left to right, $9.6\times 10^{-19}$, $9.0\times 10^{-20}$, $2.2\times 10^{-19}$ and $3.4\times 10^{-18}\,$W$\,$m$^{-2}$; for $10000\,$K they are $4.6\times 10^{-18}$, $2.6\times 10^{-19}$, $1.1\times 10^{-18}$ and $3.1\times 10^{-17}\,$W$\,$m$^{-2}$. }
\label{fig:mainplot1_Rc50}
\end{figure*}

\begin{figure*}[!ht]
\centering
\includegraphics[clip=, width=1.0\linewidth]{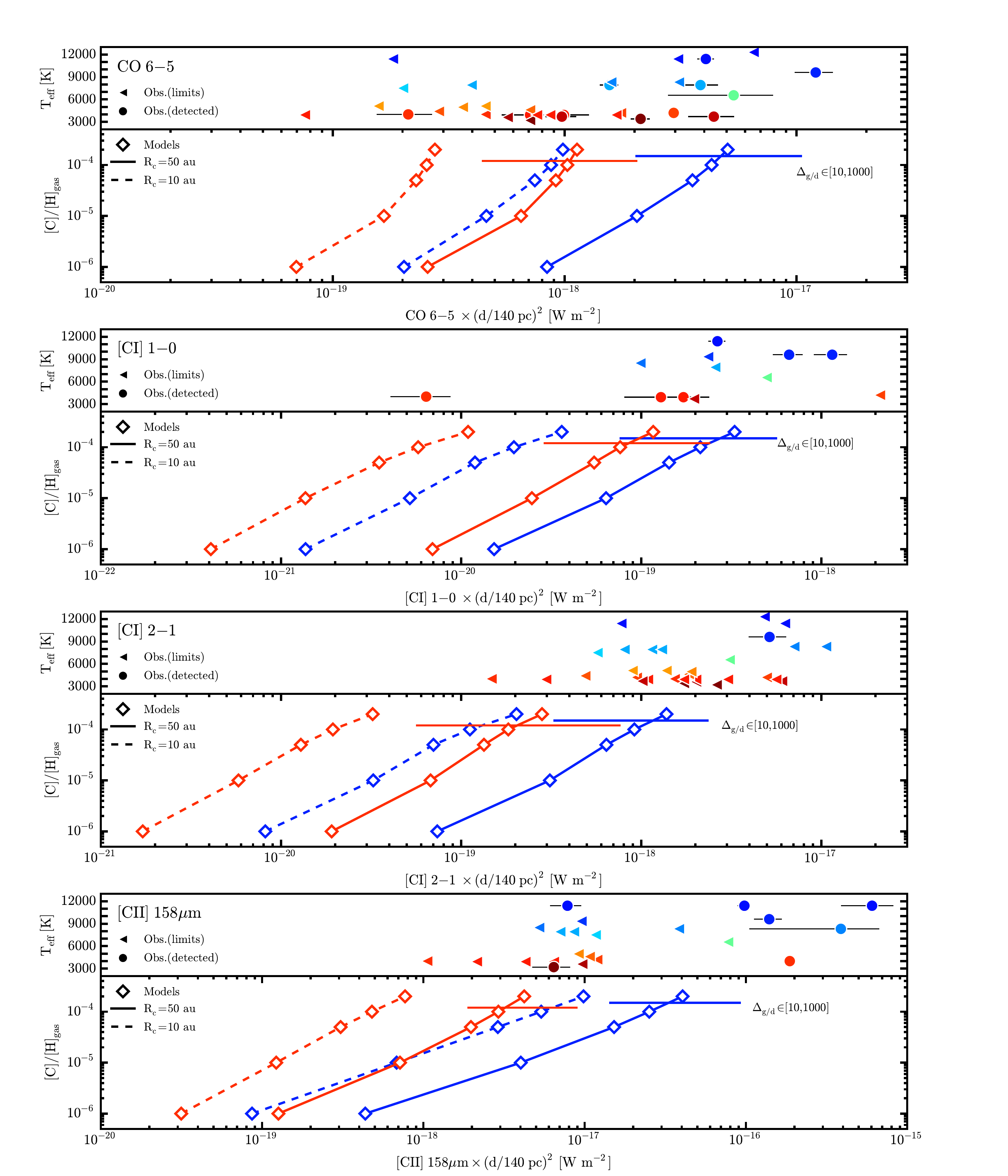}
\caption{APEX observations (circles, triangles) and DALI full disk models (diamonds) of the \cosixfive\ (top), \cionezero, \citwoone, and \ciiline\ (bottom) transitions. All values are colour-coded for the stellar \teff, with T~Tauri stars red and Herbig~Ae/Be stars blue. The fluxes are plotted with stellar \teff\ values in the upper subpanels (with \chgas~$=1.35\times 10^{-4}$ and $R_{\rm c}=50\,$au fixed for the fiducial model), and with the gas-phase total elemental carbon abundance, \chgas, in the lower subpanels (solid lines connect models with $R_{\rm c}=50\,$au, dashed lines $R_{\rm c}=10\,$au). Horizontal lines in the lower subpanels show the flux changes from varying the gas-to-dust ratio by a factor of ten around the fiducial T~Tauri and Herbig~Ae/Be models.}
\label{fig:chgasfluxes}
\end{figure*}

\subsection{The degeneracy of \chgas\ with other parameters}

In Fig.~\ref{fig:mainplot1_Rc50}, the effect of various disk model parameters on the emergent line fluxes of CO, \ci\ and \cii\ is shown, and compared with the effect of varying \chgas\ for the stars representing the T~Tauri and Herbig~Ae/Be classes. The parameters with the smallest influence on the line fluxes include the flaring parameters ($\psi$ and $h_{\rm c}$), the mass and scaleheight ratio of large to small grains ($f$ and $\chi$), the inner hole size ($r_{\rm hole}$) and the cosmic ray ionization rate ($\zeta_{\rm cr}$). Considering the \cionezero\ line flux, all these parameters individually lead to variations of a factor of no more than two within the full range of values listed in Table~\ref{tab:gridparams}. These parameter dependencies are discussed in more detail in Appendix~\ref{sec:degeneracies}.

The disk surface density scaling ($\Sigma_{\rm c}(R_{\rm c})$ for $R_{\rm c}=\mathrm{const}$) and power law index ($\gamma$) both yield up to a factor of five variations of line flux, but the corresponding change in $\Sigma_{\rm c}(R_{\rm c})$ is equivalent to two orders of magnitude in disk mass. 

As the mass of dust contributing to the optical through millimeter opacity is known to reasonable accuracy, the next main uncertainty is the gas to dust ratio (\gdrat). A change of one order of magnitude in this ratio induces a change of a factor of three in the \ci\ lines, and a factor of two in the CO and \cii. This is a comparable influence to one order of magnitude in \chgas. A gas-phase carbon underabundance of one to two orders of magnitude, such as has been proposed for TW~Hya \citep{Favreetal2013}, is easily distinguishable from all the reasonable parameter variations explored above. Furthermore, while factor of two changes in \chgas\ are easily masked by various parameter uncertainties, a carbon underabundance of one order of magnitude can be readily identified as long as $R_{\rm c}$ is determined to within a factor of two and the disk flaring type -- flat, with $\psi\approx 0.1$; or flared, with $\psi\gtrsim0.2$ -- is known. Both of these are reasonable requirements, especially as the spatial resolution of ALMA will allow to determine the radial size of disks within a few hundred parsecs to much better than a factor of two.

In summary, if the large-scale radial structure of the disk is known to a level reasonably expected in the ALMA era, and the flaring state of the disk is known (from the mid- to far-infrared SED, for example), a gas-phase carbon underabundance of an order of magnitude or more can be distinguished from reasonable variations of individual parameters. A more accurate estimate of \chgas\ will require detailled modelling of the disk of interest, to avoid unfavourable combinations of parameter uncertainties. Masking an underabundance as severe as two orders of magnitude in \chgas\ would require unreasonable fine-tuning of such parameter variations, however.

\subsection{Comparison with observations}\label{sec:comparison}

In Fig.~\ref{fig:chgasfluxes}, we compare the APEX observations of  \cosixfive, \cionezero, and \citwoone with a grid of DALI models. We also show the literature \cii\ fluxes. To cover the spectral types in our sample, we vary the stellar properties from a \teff~$4000\,$K T~Tauri star with UV excess to a $12000\,$K Herbig star, in steps of $2000\,$K. The luminosities are given in Table~\ref{tab:uvlums}. For the \teff~$=4000\,$K and $10000\,$K stars, we show two disk sizes, $R_{\rm c}=10$ and $50\,$au, and vary the gas-phase carbon abundance, \chgas~$\in{[10^{-6},2\times 10^{-4}]}$. The total disk mass is fixed at $M_{\rm disk}=10^{-2}\,$\msol, except when varying the gas-to-dust ratio by a factor of ten, where we keep the dust mass fixed. All fluxes are normalized to a distance of $140\,$pc, and colour-coded for the stellar effective temperature.

As detailed in the following two subsections, most of our observations fall within reasonable variations of the model parameters. Due to limited sensitivity, none of the observed lines strongly constrains the carbon abundance in most sources.  A subset of sources are brighter than the models at some of the targeted frequencies. These are very flared disks with large inner cavities (HD~142527, HD~97048, Oph~IRS~48) or with a powerful jet (\cii\ towards DG~Tau).

\subsubsection{T~Tauri systems (spectral types G, K, M)}

\textbf{\cosixfive: }The observations almost entirely fall between the fiducial $R_{\rm c}=10$ and $50\,$au models with an ISM-like \chgas. This underlines the importance of knowing the radial extent of the gas disk for constraining \chgas. The three disks with the deepest CO limits (BP~Tau, T~Cha) or detections (TW~Hya) may be carbon-depleted, while the three disks with the brightest detections (AS~205, Sz~33, V806~Tau) are likely contaminated by envelope or disk wind emission.

\textbf{\cionezero: }The detection towards TW~Hya confirms its low \chgas\ ratio, while the other four observations lie above most of the models. The two detections among these four are likely residual envelope or molecular cloud emission, as the line profiles are very narrow. Narrow lines are unexpected for both AA~Tau, which is a well-known edge-on disk, as well as for DM~Tau, which has $i{=}32^{\circ}$, and where the \cosixfive\ line is substantially broader than the \cionezero\ (see Figure~\ref{fig:dmtau}).

\textbf{\citwoone: }All targeted disks with the exception of TW~Hya have upper limits above the model grid. The upper limit for TW~Hya does not strongly constrain \chgas, but is consistent with the low value inferred from \cosixfive\ and \cionezero.

\textbf{\ciiline: }The upper limits from \citet{Howardetal2013} are generally not yet sensitive enough to lie in the model parameter space, with the exception of the datapoint for TW~Hya from \citet{Thietal2010}. The second-lowest \ciiline\ upper limit is obtained for AA~Tau. The point lies close to an ISM-like model \chgas\ and does not provide a strong constraint.

\textbf{Additional comments: }BP~Tau and T~Cha have distance-normalized \cosixfive\ upper limits below the detection of TW~Hya, which is an extremely weak CO, \ci\ and \cii\ emitter. Combined with the known gas disk radius of ${\approx}215\,$au \citep{Andrewsetal2012}, the CO and \ci\ line fluxes for TW~Hya are consistent with a factor of $10$--$100$ underabundance of gas-phase carbon \citep[][Kama et al. submitted]{Favreetal2013}. The gas disk of BP~Tau extends to ${\approx}100\,$au, and the CO-based disk mass is only $M_{\rm disk}=1.2\times 10^{-3}$ \citep{Dutreyetal2003}. The gas disk of T~Cha extends to ${\approx}200\,$au \citep{Huelamoetal2015}. Given their considerable radial sizes, comparable to that of TW~Hya, the disks of both BP~Tau and T~Cha thus have either a low \gdrat\ or a low \chgas\ ratio. DM~Tau and IM~Lup are large disks with \cosixfive\ fluxes comparable to our models with interstellar \chgas. GG~Tau~A appears similar but is a special case, as it is a large circum-triple ring -- not directly comparable to any of our models \citep{Guilloteauetal1999, DiFolcoetal2014}. AS~205, Sz~33 and V806~Tau have substantially larger \cosixfive\ fluxes than predicted by our fiducial models, placing them close to the Herbig disks. As discussed earlier, remnant envelopes and disk winds could dominate emission in these systems. If the systems are very young, the stellar luminosity for a given \teff\ could also be substantially larger.

\subsubsection{Herbig~Ae/Be systems (spectral types B, A, F)}

\textbf{\cosixfive: }All non-extended, non-contaminated detections of CO around early-type stars are from group~I systems, in the \citet{Meeusetal2001} classification of Herbig~Ae/Be disks as flaring/warm (group I) and flat/cold (group II). This is consistent with the flaring disks being warmer and thus stronger emitters (see also the effects of $\psi$ and $h_{\rm C}$ in Fig.~\ref{fig:mainplot1_Rc50}). All the detected disks have inner holes -- HD~97048, HD~100546 and HD~169142 \citep[hole radii $34$, $13$, and $23\,$au, respectively,][]{Maaskantetal2013, Panicetal2014, Walshetal2014b}. However, the line is not detected towards HD~139614 \citep[$5.6\,$au][]{ Matteretal2014}, which has a dust mass and spectral type very similar to HD~169142 \citep[$\approx 1\times 10^{-4}$~\msol\ and A8/A5][]{Dentetal2005, Panicetal2008, Maaskantetal2013}. This, combined with its large outer radius of $150\,$au \citep{Matteretal2015}, suggests a low \gdrat\ or \chgas. Two early-type disks lie below even the late-type locus on the \cosixfive\ axis. Two of these resemble debris disks: HD~141569 is a transitional disk with no substantial surface density at radii below $\sim95$~au \citep{Dentetal2005, Jonkheidetal2006}, and HD~100453 has a CO-based gas mass of only $10^{-4}\,$\msol\ and an inner cavity of $20\,$au \citep{Khalafinejadetal2015, Collinsetal2009}.

\textbf{\cionezero: }Detections are obtained towards three flaring, group~I disks. The detection in HD~100546 shows a clear double-peaked Keplerian profile, while the other two (HD~97048 and Oph~IRS~48) are likely contaminated by foreground or extended emission. The upper limits towards HD~163296 and HD~169142 allow both small and large disk models with an ISM-like \chgas, while the upper limit towards HD~104237 lies between the $R_{\rm c}=10$ and $50\,$au models.

\textbf{\citwoone: }The only detection is towards the very flared and embedded disk, HD~97048. Nearly all the upper limits lie close to, or above, the large ($R_{\rm c}=50\,$au) disk models with an ISM-like \chgas. The deepest distance-normalized upper limit is obtained for HD~100453, a gas- or carbon-depleted transitional disk described earlier in this subsection.

\textbf{\ciiline: }Similarly to \cosixfive, all detections of this line are from group~I disks, alhtough contamination cannot be checked as easily as for CO. None of the \emph{Herschel}/PACS \ciiline\ upper limits from \citet{Meeusetal2012, Dentetal2013, Fedeleetal2013a} on our Herbig~Ae/Be targets provide strong constraints on the gas content or carbon abundance. Most of the upper limits lie between the fiducial $R_{\rm c}=10$ and $50\,$au models with an ISM-like \chgas. We have also compared our model grid with the full sample of the above \emph{Herschel} studies (not plotted), and find that the only anomalously deep upper limits are for debris disks, where the gas content is low (e.g., HR~1998, 49~Cet, HD 158352).

\textbf{Additional comments: }HD~50138 may be an evolved star. It was included in the sample as a possible protoplanetary disk and is often treated as such in the literature, but see e.g., \citet{Ellerbroeketal2015} for a discussion of its unclear nature. Its \cosixfive\ upper limit falls between our $R_{\rm c}=10$ and $50\,$au models, while the \cii\ detection is an order of magnitude brighter than any of our Herbig~Ae/Be disk models.

\subsection{Summary of carbon abundance constraints}

Our APEX survey observations of the \cionezero\ and \citwoone\ lines, with typical detection limits within a factor of a few of $10^{-18}\,$W$\,$m$^{-2}$ at $140\,$pc, are not yet sensitive enough to strongly constrain \chgas\ in most disks. They require a factor of three to ten improvement. This corresponds to observations of ${\approx}10\,$hours per source with APEX, if overheads are included. Our deep exposures towards HD~100546 and TW~Hya demonstrate that detections can be made in this way in the brightest disks. On ALMA, a similar sensitivity can be reached in ${\approx}1\,$h for a synthesized beam of ${\approx}0.5\,''$. ALMA is essential for extending the \ci\ detection sample beyond the few nearest, brightest disks. Equally importantly, any extended emission around the disk will be filtered out by ALMA. Its spatial resolution is sufficient to resolve the CO snowline and study any associated gas-phase carbon abundance variations in many systems, including TW~Hya.

The sensitive detection of \cionezero\ towards TW~Hya (corresponding to $7\times10^{-20}\,$W$\,$m$^{-2}$ at $140\,$pc; note that the source is at $55\,$pc) is consistent with a factor of $100$ underabundance of gas-phase carbon, confirming the result of \citet{Favreetal2013}. For HD~100546, the detection ($2\times10^{-19}\,$W$\,$m$^{-2}$ at $140\,$pc; the source is at $97\,$pc) suggests a gas-phase abundance close to the interstellar one, consistent with earlier work by \citet{Brudereretal2012}. For both sources, detailed models of the disk structure, in particular constraining the radial extent of the disk, are needed to determine \chgas\ with better precision. We present such modelling in a companion paper (Kama et al. submitted).

Several sources in our sample have relatively deep upper limits on CO or \ci. They could be either gas- or carbon-poor. These systems are BP~Tau, T~Cha, HD~139614, HD~141569, and HD~100453. The latter two have low CO-based gas mass estimates and are often considered debris disk like systems, however without an absolute gas mass determination a carbon depletion cannot be entirely ruled out.

\section{Conclusions}\label{sec:conclusions}

We present observations and modelling of the main gas-phase carbon reservoirs in protoplanetary disk atmospheres, \catom\ and CO. We observed $37$ disks with the APEX telescope, and employed DALI physical-chemical models to interpret the data and investigate the relation of CO, \ci\ and \cii\ emission with the gas-phase carbon abundance, \chgas.

\begin{enumerate}
\item{Among our full sample, \cosixfive\ is detected (observed) towards $13$ ($33$) sources; \cionezero\ in $6$ ($12$); and \citwoone\ in $1$ ($33$). The \ci\ detections are extended or of unclear origin in all sources except TW~Hya and HD~100546.}
\item{We detect \cionezero\ from the disks around TW~Hya and HD~100546, the first unambiguous detections of this line in protoplanetary disks. The HD~100546 emission has a symmetric double-peaked line profile.}
\item{Based on a grid of models, we find that the survey sensitivity, typically $10^{-19}\,$W$\,$m$^{-2}$ for \cionezero\ and $10^{-18}\,$W$\,$m$^{-2}$ for \citwoone, needs to be improved by a factor of at least three to obtain useful constraints on \chgas\ for most systems.}
\item{An underabundance of one order of magnitude for gas-phase carbon cannot easily be masked by other disk properties, if they are known to a level made possible by ALMA, \emph{Herschel} and \emph{Spitzer} (Fig.~\ref{fig:mainplot1_Rc50}). Larger underabundances are even more reliably identified. The most essential information for constraining the gas-phase carbon abundance is the radial extent of the gas disk.}
\item{A comparison with our general grid of models suggests that gas-phase carbon is underabundant by up to a factor $100$ in TW~Hya, while for HD~100546 the comparison suggests a roughly interstellar abundance. For a detailed analysis, see Kama et al. (submitted).}
\item{BP~Tau, T~Cha, HD~139614, HD~141569, and HD~100453 warrant follow-up as potentially carbon-depleted disks.}
\item{The low detection rate of \cii\ emission with \emph{Herschel}/PACS in T~Tauri disks is due to insufficient sensitivity, while the Herbig~Ae/Be systems with no detection are typically gas-poor debris disks.}
\end{enumerate}

\begin{acknowledgements}
We thank the anonymous referee for constructive comments which helped to improve the paper, Arnaud Belloche and the APEX staff for assistance during the observations, and Matthijs van der Wiel for discussing his SPIRE observations with us. This work is supported by a Royal Netherlands Academy of Arts and Sciences (KNAW) professor prize, the Netherlands Research School for Astronomy (NOVA), and by the European Union A-ERC grant 291141 CHEMPLAN. A.K. acknowledges support from the Foundation for Polish Science (FNP) and the Polish National Science Center grant 2013/11/N/ST9/00400. R.J.W. is supported by NASA through the Einstein Postdoctoral grant number PF2-130104 awarded by the Chandra X-ray Center, which is operated by the Smithsonian Astrophysical Observatory for NASA under contract NAS8-03060. This publication is based on data acquired with the Atacama Pathfinder Experiment (APEX). APEX is a collaboration between the Max-Planck-Institut f\"{u}r Radioastronomie, the European Southern Observatory, and the Onsala Space Observatory. CHAMP$^+$ was constructed with support from the Netherlands Organization for Scientific Research (NWO), grant 600.063.310.10.
\end{acknowledgements}

\bibliographystyle{aa}
\bibliography{disks}

\begin{appendix}

\section{Structure of the disk models}\label{sec:modeldetails}

\begin{figure}[!ht]
\centering
\includegraphics[clip=, width=1.0\columnwidth]{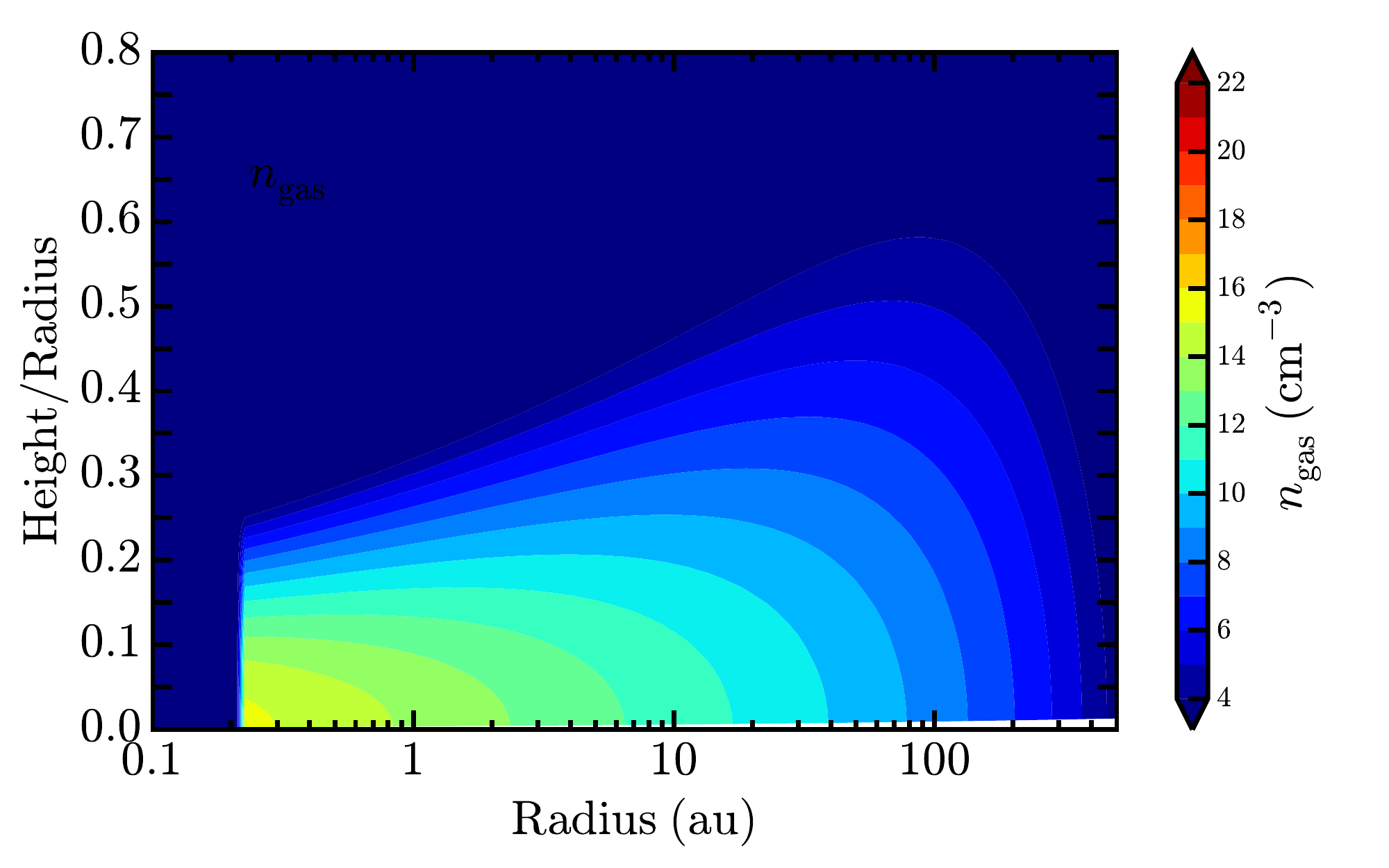}
\caption{The gas density structure of the fiducial disk model.}
\label{fig:ngas}
\end{figure}

\begin{figure*}[!ht]
\centering
\includegraphics[clip=, width=0.9\linewidth]{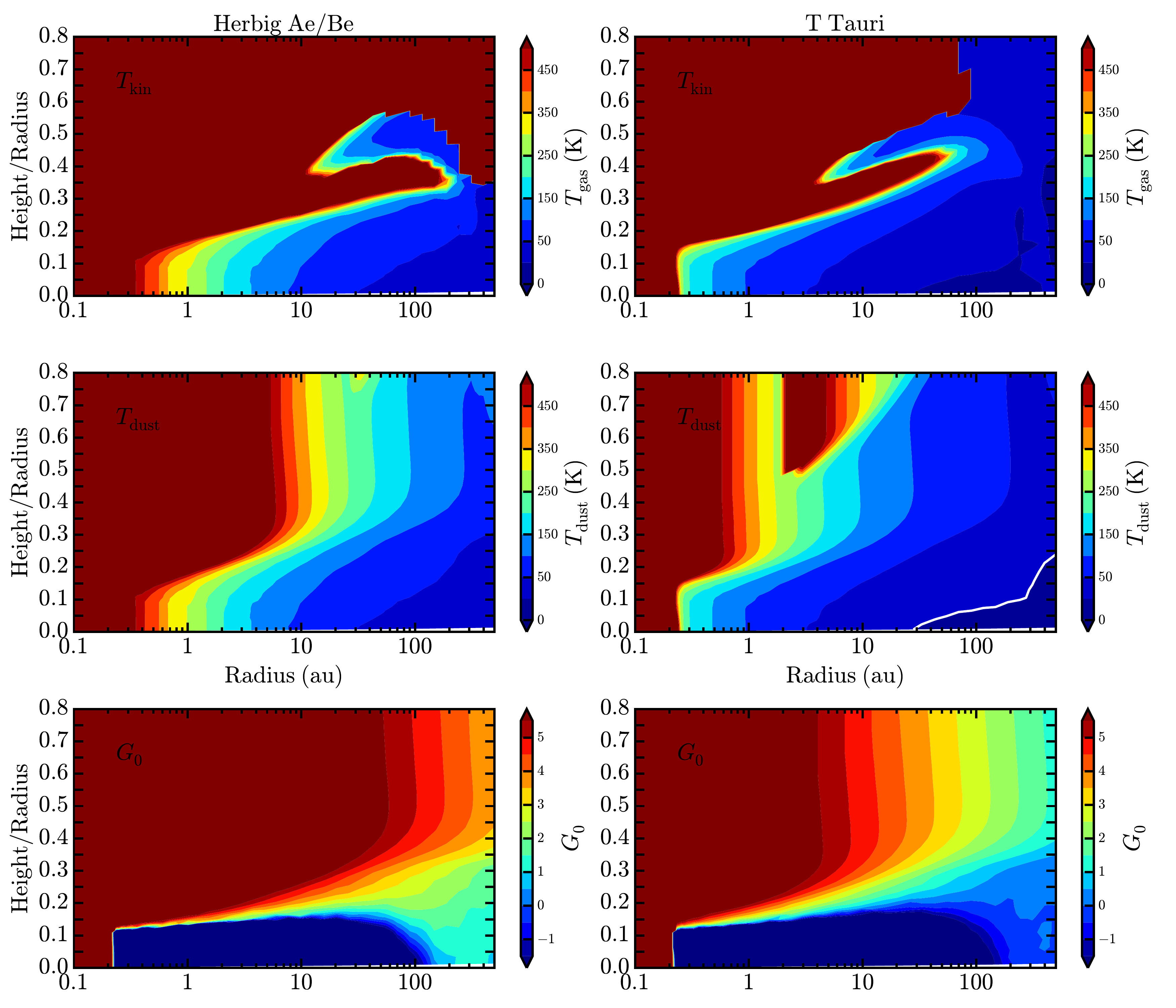}
\caption{The fiducial disk models for a \teff~$10000\,$K star (Herbig~Ae/Be, left column) and a \teff~$4000\,$K star with an ultraviolet excess (representing T~Tauri disks, right column). From top to bottom, the panels show the gas and dust temperature, and the ultraviolet field in standard interstellar field units, G$_{0}$.}
\label{fig:diskstruct}
\end{figure*}

In Figs.~\ref{fig:ngas} and \ref{fig:diskstruct}, we show the density structure and the gas and dust temperature, as well as the ultraviolet field in interstellar units (denoted G$_{0}$) for the fiducial T~Tauri and Herbig~Ae/Be disk models.

\section{The degeneracy of \chgas\ with other parameters}\label{sec:degeneracies}

\textbf{The elemental gas-phase carbon abundance (\chgas). }Order-of-magnitude changes in \chgas\ outweigh the impact of reasonable variations of any other individual parameter. The fluxes of all lines except \cosixfive\ change by a factor of five for a factor of ten change in \chgas, implying that large underabundances of the magnitude proposed for TW~Hya -- more than one order of magnitude -- can be reliably identified, if the disk structure is reasonably constrained.

\textbf{Gas to dust ratio (\gdrat). }This parameter is varied by changing the gas mass. Assuming a line becomes optically thick at some fixed column density of gas, and ignoring other effects, the emitting area scales as $A_{\rm em}\propto \Delta_{\rm g/d}^{2/\gamma^{\prime}}$, where $\gamma^{\prime}$ is the the power-law index of the local surface density profile, $\Sigma(r)\propto r^{-\gamma^{\prime}}$. For the global $\gamma=1.0$, the CO emission thus scales weakly with the gas mass in the exponentially decreasing outer tail of the surface density (large local $\gamma^{\prime}$). Once the optically thick surface is limited to smaller radii, the relation becomes steeper (small local $\gamma^{\prime}$). The initial slow decrease of the line emission with gas mass is also related to the smaller importance of the cold outer disk, which can be interpreted as a further decline of the surface density of particles in the required upper energy state. For \catom\ line fluxes, one order of magnitude in \gdrat\ ratio gives the same change as a factor of three in \chgas. Therefore, an order-of-magnitude scale underabundance of gas-phase carbon can be identified even with a poorly known \gdrat.

\textbf{The disk surface density power law index ($\gamma$). }If $\Sigma_{\rm c}$ and $R_{\rm c}$ are fixed, increasing gamma increases the fluxes of all lines. From $\gamma=0.8$ to $1.5$, the \catom\ line fluxes can increase by a factor of three to ten. This is related to an increase in disk mass, which doubles. Note that, in the power-law surface density with an exponential taper, $\gamma=1.5$ yields quite a shallow surface density profile in the exponential tail, which is reflected in an increased CO emitting area in the outer disk.

\textbf{Disk surface density anchorpoint ($R_{\rm c}$). }Given a $\gamma$ and $\Sigma_{\rm c}$, this parameter sets the radial size of the disk, which can be defined as the extent of optically thick CO emission.

\textbf{The disk surface density scaling ($\Sigma_{\rm c}(R_{\rm c})$). }For a fixed $R_{\rm c}$, this parameter varies the disk mass. For a range of three orders of magnitude in gas mass, the largest line flux variation amplitude (\cosixfive) is a factor of ten.

\textbf{The flaring parameters ($\psi$ and $h_{\rm c}$). }The variation amplitude of $\psi$ leads to up to a factor of five in CO and \ci\ line fluxes, but \cii\ can vary by an order of magnitude. The CO and \catom\ line fluxes depend on $\psi$ in all cases, while \cplus\ becomes more sensitive with increasing ultraviolet flux ($10000$~K emission). For a pure $4000$~K radiation field, \tkin\ is too low in the tenuous upper layers of the disk to produce substantial \ciiline\ emission. Varying $h_{\rm c}$ has up to a factor of two impact on the lines, as it changes the volume of warm gas.

\textbf{Mass ratio of large to small grains ($f$). }Decreasing the fraction of large grains increases the optical depth of the disk material at short wavelengths, making the emitting regions colder. The ionized carbon is in completely optically thin gas and is less affected. From $f = 0.2$ to $0.999$, the variation in CO line flux does not exceed a factor of three, with \catom\ and \cplus\ less affected.

\textbf{Inner hole size ($r_{\rm hole}$). }An inner hole of up to $100\,$au in size has only a factor of $\leq 2$ impact on any of the line fluxes. This parameter is only important if the disk is very small.

\end{appendix}

\listofobjects
   
\end{document}